\documentclass[aps,pre,superscriptaddress,showpacs,nofootinbib]{revtex4}
\usepackage{amsfonts,amssymb,amsmath,bm,calligra,euscript,mathrsfs}
\usepackage[utf8]{inputenc}
\usepackage[T1]{fontenc}
\usepackage{color,soul}
\usepackage{textcomp}
\usepackage{graphicx}
\usepackage[tight,sf]{subfigure}
\usepackage{hyperref}
\newcommand{\sech}{\mathrm{sech} \,}
\newcommand{\arcsinh}{\mathrm{arcsinh} \,}
\newcommand{\am}{\mathrm{am}}
\newcommand{\sn}{\mathrm{sn}}
\newcommand{\cn}{\mathrm{cn}}
\newcommand{\dn}{\mathrm{dn}}

\begin{document}

\title{Cylindrical  confinement of semiflexible polymers}

\author{Pablo V\'azquez-Montejo}
\email[]{pablov@andrew.cmu.edu}
\affiliation{Department of Physics, Carnegie Mellon University,\\5000 Forbes
Ave, Pittsburgh, PA 15213, USA}
\author{Zachary McDargh}
\email[]{zmcdargh@andrew.cmu.edu}
\affiliation{Department of Physics, Carnegie Mellon University,\\5000 Forbes
Ave, Pittsburgh, PA 15213, USA}
\author{Markus Deserno}
\email[]{deserno@andrew.cmu.edu}
\affiliation{Department of Physics, Carnegie Mellon University,\\5000 Forbes
Ave, Pittsburgh, PA 15213, USA}
\author{Jemal Guven}
\email[]{jemal@nucleares.unam.mx}
\affiliation{Instituto de Ciencias Nucleares, Universidad Nacional Aut\'onoma
de M\'exico \\ Apdo. Postal 70-543, 04510 M\'exico, DF, M\'EXICO}

\begin{abstract}
Equilibrium states of a closed semiflexible polymer binding to a cylinder are described. This may be either by confinement or by constriction. Closed completely bound states are labeled by two integers: the number of oscillations, $n$, and the number of times it winds the cylinder, $p$, the latter being a topological invariant. We examine the behavior of these states as the length of the loop is increased by evaluating the energy, the conserved axial torque and the contact force. The ground state for a given $p$ is the state with $n=1$; a short loop with $p=1$ is an elliptic deformation of a parallel circle; as its length increases it elongates along the cylinder axis, with two hairpin ends. Excited states with $n \geq 2$ and $p=1$ possess $n$-fold axial symmetry. Short (long) loops possess energies $\approx p E_0$ ($nE_0$), with $E_0$ the energy of a circular loop with same radius as the cylinder; in long loops the axial torque vanishes. Confined bound excited states are initially unstable; however, above a critical length each $n$-fold state becomes stable: the folded hairpin cannot be unfolded. The ground state for each $p$ is also initially unstable with respect to deformations rotating the loop off the surface into the interior. A closed planar elastic curve aligned along the cylinder axis making contact with the cylinder on its two sides is identified as the ground state of a confined loop. Exterior bound states behave very differently, if free to unbind, as signaled by the reversal in the sign of the contact force. If $p=1$, all such states are unstable. If $p \geq 2$, however, a topological obstruction to complete unbinding exists. If the loop is short, the bound state with $p=2$ and $n=1$ provides a stable constriction of the cylinder, partially unbinding as the length is increased. This motif could be relevant to an understanding of the process of membrane fission mediated by dynamin rings.
\end{abstract}


\maketitle

\section{Introduction}

\begin{figure}[htb] \begin{center} \begin{tabular}{ccc}
$\vcenter{\hbox{\includegraphics[scale=0.45]{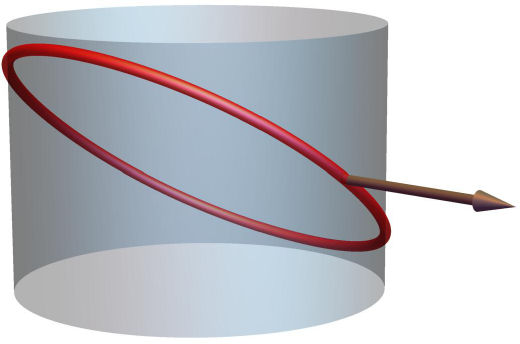}}}$ &
$\vcenter{\hbox{\includegraphics[scale=0.5]{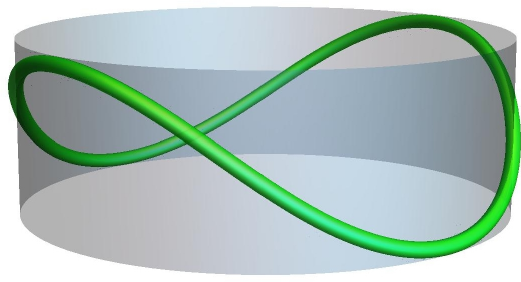}}}$ &
$\vcenter{\hbox{\includegraphics[scale=0.4]{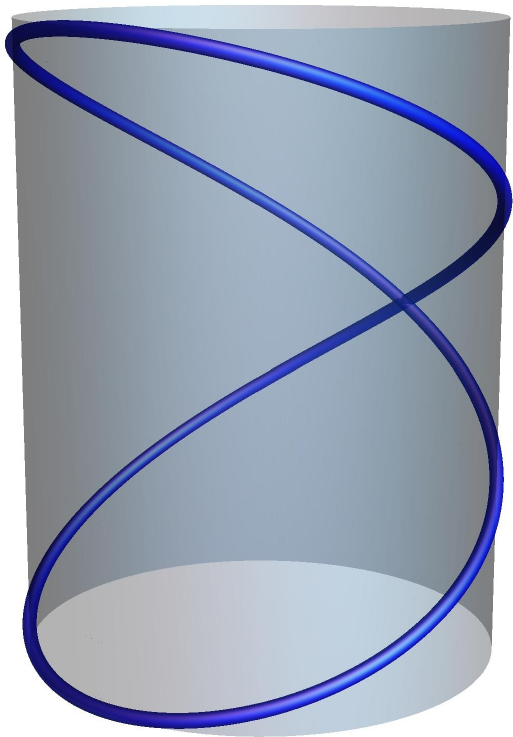}}}$ \\ {\small (a)} &
{\small (b)} & {\small (c)} \end{tabular} \end{center} \caption{(Color online)
(a) Elliptic state with $n=1$ and $p=1$. (b) Two-fold state with $n=2$ and
$p=1$. (c) Folded figure-eight state, $n=1$ and $p=2$. The arrow indicates the
local normal vector to the surface, which signifies the direction along which confining forces have to
act in order to keep the polymer on the surface, with a sign convention further
discussed in the text.} \label{Fig1}
\end{figure}

The confinement of semi-flexible polymers or filaments by curved surfaces is
increasingly recognized to be a key element in a number of physical processes.
The best studied examples are presented in biology: DNA, actin filaments, and
microtubules, for instance, have persistence lengths of $50\,{\rm nm}$
\cite{Marko1995}, $\sim\!16\,\mu{\rm m}$  \cite{Ott1993} and several
millimeters \cite{Gittes1993}, respectively. Because such polymers need to
navigate an intracellular environment crowded by membranes and proteins, over
and over again one needs to understand how they interact with these structures.
DNA in eukaryotic cells, for instance, makes a virtue of necessity, wrapping
around the core of cylindrical histones to facilitate their condensation
\cite{Luger1997, Schiessel2003, Cherstvy2014}; intermediate filaments (such as
spectrins or lamins) may adsorb onto membranes and, in the process, modify
their elastic properties \cite{Discher1994, Dechat2008, Dahl2008}; membrane
fission is also frequently driven by the polymerization of tightly winding
stiff filaments of the protein dynamin, whose winding around the neck of a
nascent vesicle is believed to cut the vesicle from its parent membrane
\cite{Morlot2013, Kozlov2001, Hinshaw1994}. Since in these and
other examples the persistence length can be quite a bit larger than the radii
of curvature to which these polymers are confined, it is a good approximation
to ignore additional thermal fluctuations and focus on ground state solutions,
which is what we do in this paper. The opposite limit, confining highly
flexible filaments, is a classical topic in polymer science, which is reviewed
in Ref.~\cite{DeGennes1999}. While stiff filaments are especially frequent and well studied in the context of biology, they also occur in other situations,
and confinement issues arise there, too. For instance, single walled carbon
nanotubes have a diameter dependent persistence length in the tens of
micrometer range \cite{Fakhri2009}, and their adsorption onto surfaces or their interplay with other mesoscopic microstructures is currently a subject of
considerable interest for engineers as well as physicists \cite{Thostenson2001, Wei2007}.
\vskip1pc \noindent
The defining feature of semi-flexibility is that the polymer's persistence length substantially exceeds molecular scales, such as the monomer size or thickness. This is physically important, because it means that the energetics of bending decouples from the minutiae of the chemical structure and can thus be described to very good accuracy by a continuum-elastic Hamiltonian depending largely on geometry. The simplest such energy functional describes a semi-flexible polymer as a one-dimensional smooth space curve and quadratically penalizes its Frenet curvature, $\kappa(s)$ \cite{Kratky1949, Langer1984, Kierfeld2008, SingerSantiago2008},
\begin{equation} \label{Hamk2} H_{\rm B} = \frac{A}{2}\, \int {\rm d}s \;
\kappa(s)^2 \ ,
\end{equation}
where $s$ is the arc length along the curve and ${A}$ is the bending rigidity. Since there are no other energy terms competing
with the bending, from now on we set the bending rigidity to unity, $A=1$. A
functional variation $\delta H_{\rm B}=0$ leads to the Euler-Lagrange (EL)
equations which characterize the solutions that minimize this energy.
\vskip1pc \noindent
Substantially more complicated functionals are not only conceivable
but have also  been studied, adding, for instance, stretching, twisting
\cite{Love, Antman, LandauElasticity1986}, spontaneous curvature
\cite{AudolyBook} or the higher derivative Frenet torsion
\cite{CapoChryssGuv2002}. An environmental bias may imply a separate dependence on the geodesic and normal curvatures, or even the geodesic torsion
\cite{GuvValVaz2014}. For the purpose of this paper we restrict our attention
to the elementary energy Eq. (\ref{Hamk2}). The problem is still non-trivial
because the energy needs to be amended with the constraint confining the
space-curve to the surface. One needs to accommodate this constraint in the
calculus of variations, as described, for example, in \cite{GuvVaz2012}.  We only consider the limiting case in which the surface is much stiffer than the
polymer itself, so that the shape of the surface remains unaffected. The
opposite limit, in which the polymer is infinitely rigid and the surface
adjusts, has recently been studied by Bo\v{z}i\v{c} \emph{et al.}
\cite{Bozic2014}, who considered a circular ring constricting an axisymmetric
neck. Of course, in many physically relevant situations both the polymer and
the surface respond to each other's presence. However, the discussion of two
elastic objects of different dimensionalities pitting their forces against each other through a mutually confining geometric constraint is a long story which will have to await a future treatment.
\vskip1pc \noindent
The first general discussion of surface-constrained elastic curves \`a la Eq. (\ref{Hamk2}) was given by Nickerson and Manning \cite{Manning1987, Nickerson1988}, who derived the EL equation for this constrained minimization problem. As an application, they studied cylindrical confinement with a particular choice of boundary conditions \cite{Nickerson1988}. Later, Marky and Manning considered the wrapping of DNA around a histone octamer, taking into account an adhesive interaction between the polymer and its substrate \cite{Marky1991}.
More recently, the problem was treated by van der Heijden, who also included a
twist degree of freedom on isotropic  \cite{vdH2001} and anisotropic
\cite{vdH2002} rods with forces applied at the boundaries. Later still, van der Heijden \emph{et al.} \cite{vdH2006} studied self-contacts of a cylindrically
confined elastic rod. In this work, the elastic curve is treated as a
Kirchhoff rod or more generally within the context of Cosserat theory. We follow the strategy introduced by two of the authors in Ref. \cite{GuvVaz2012}, where the surface constraint was enforced using a local Lagrange multiplier in
the variational principle, which can be identified as the external source of
stresses to which the polymer is subject due to the confinement. This provides
direct access to the magnitude of the confining forces, as well as their sign,
which will vary along the polymer. Formally, the multiplier is identified with
the normal projection of the derivative of the force vector, which permits one
to quantify the loss of Euclidean invariance of the constrained system. In this approach it is unnecessary to assume any constitutive relations among the
bending moments, the energy density, and the strains: the balance of forces and torques follow from the minimization of the energy; these two vectors are also
identified. For simplicity we restrict our attention to closed polymer
loops.
\vskip1pc \noindent
The equilibrium shapes of semiflexible polymer loops, modeled as elastic curves, confined by a sphere were examined in some detail in Ref. \cite{GuvVaz2012} (see Refs. \cite{Stoop2011, Najafi2012} for a more physical and realistic treatment, involving numerical analysis, see also
\cite{Shin2014, Shin2015} for a detailed study of the dynamics
of confined semiflexible polymers employing simulations, as well as
\cite{Cherstvy2011, Carvalho2015} addressing the effect of electrostatics on
the adsorption of polymers rings onto surfaces with spherical and cylindrical
geometry). The analytical treatment in Ref. \cite{GuvVaz2012} was facilitated
by identifying the three conserved quantities---corresponding to the three
components of the torque---associated with the spherical symmetry. The EL
equations could be integrated completely in terms of elliptic functions. It was shown that the physics of these states depends very sensitively on the length
of the loop.  Indeed, the ground state itself alternated between two states as
the length was increased.  The forces transmitted to the sphere was also shown
to depend both non-monotonically and discontinuously on this length. As the
length becomes increasingly large, all bound states tend increasingly towards
geodesic behavior. The aim of the present paper is to investigate
systematically the binding of a semi-flexible polymer loop to a cylinder, not
only by  its confinement within it but also by the forced constriction of the
cylinder by an exterior loop. In the absence of additional adhesive forces the
latter has no analogue on a sphere.
\vskip1pc \noindent
A cylinder is distinguished from a sphere both geometrically (curvature anisotropy) and topologically. Specifically, it is non-compact. Each of these features will play a role: some times alone, other times together with unexpected outcomes. We begin by showing how the conservation laws associated with the residual Euclidean symmetries conspire not only  to yield a quadrature, but also to provide the appropriate observables for characterizing the underlying physics. While we lose rotational symmetry about the local surface normal, the underlying axial and translational symmetries still provide two conservation laws, one  for the axial torque and another for the axial force. These first integrals are each of third order in derivatives of the embedding functions. However, the two can be combined into a single equation eliminating one derivative, which can be cast as a simple quadrature in terms of the tangent angle. This quadrature involves three parameters: the two conserved quantities as well as a Lagrange multiplier associated with the fixed length of the polymer.
\vskip1pc \noindent
We are particularly interested in how the equilibrium states, and in particular candidate ground states, depend on the length of the loop. For a given length, such loops are characterized uniquely by two integers, the number of times the loop wraps around the cylinder, $p$, and the number of oscillations it executes in the process (its dihedral symmetry of order $n$). These states differ from their spherical counterparts, also labeled by two integers, in a number of ways. If its length is small, the loop does not yet explore the non-compact direction; the reduced symmetry however identifies the bound ground state as an elliptical deformation of a circular loop (with $n=1$, see Fig. \ref{Fig1}(a)), a state without any equilibrium counterpart on a sphere. As the length increases the loop elongates along the cylinder axis, terminating in two hairpin ends. At some point, as we show, the loop self-intersects, thereafter  scissoring--in accordion fashion--along the axial direction. States with higher dihedral symmetries are unstable initially (a state with $n=2$ is illustrated in Fig. \ref{Fig1}(b)). As the length is increased, however, these excited states also grow along the non-compact direction, forming  folded hairpins. Once they develop, it costs energy to undo these folds. The important point is that beyond some critical length, the folded state become snagged energetically on the cylinder; they are stabilized. There is no analogous process for loops binding to a sphere. Also, unlike a sphere where states morph periodically into multiple coverings of a geodesic circle with increasing loop length,  the hairpins that develop diverge increasingly from geodesic behavior, while at the same time tending towards a simple characteristic structure. Indeed, the asymptotic energy assumes the remarkably simple form, $E_{p,n} \to n E_0$, independent of $p$, where $E_0$ is the energy of a circular loop with a radius equal to that of the cylinder. It is independent of the topology, so that the energy of these limit states is highly degenerate. We describe in detail how the axial torque, the energy and, perhaps most importantly, the forces responsible for confinement depend on the length of the loop for each pair of integers, $n$ and$p$.
The direction of the normal force transmitted to the surface obviously depends
on which side the polymer is bound to: when the loop is inside, positive
(negative) values represents outward (inward) forces. If the loop is outside,
these sign conventions are reversed. The latter are especially interesting
because, as we show, an external polymer may constrict the cylinder, a
happy conspiracy permitted by the topology, an impossibility on a sphere. While in the spherical case the confining force  on an interior bound loop always
pushes it up against the sphere, we see that for cylinders the sign of the
confining force generally changes along equilibrium bound states. This would
indicate that additional forces are required to stabilize the state (deriving,
for instance, from an adhesion energy)  if it is free to unbind. In the case of interior bound states, we find that the offending regions are extremely
localized in the neighborhood of hairpins, suggesting that the state is
stabilized if the loop simply takes a shortcut off the surface so as to avoid
these regions. If the reader feels uncomfortable taking shortcuts, they may
assume that an additional constant sticky force is provided where necessary. We also provide an exact description of states confined within the cylinder
making partial contact with its walls. The corresponding ground state is a
planar loop aligned along the cylinder axis making contact on its two sides: a
vertical safety pin.  This state has lower energy than its bound counterpart.
In the absence of additional adhesive forces,  the bound interior ground
state--if sufficiently short--is expected to be unstable with respect to
deformation into the cylinder rotating it into this state. As its length
increases, however, it would appear to turn stable with respect to such
deformations, protected by  an ever-increasing energy barrier.  A bound
exterior state winding the cylinder once ($p=1$) would, on the other hand,
simply peel off relaxing towards a circular loop. If the loop winds twice or
more times around the cylinder (see Fig. \ref{Fig1}(c)), however,  this is no
longer the case.\footnote{Note that a short interior bound figure-eight
would be expected to unwind into the confined planar vertical hairpin state} A
figure-eight possesses a lower energy than a free doubly-wound circular
loop, and would be expected to remain completely attached to the cylinder if
the loop is sufficiently short. Longer loops will not be completely attached
but they would nevertheless be expected to embrace the cylinder at least
partially, applying a force at the points where they remain in contact. This is because the presence of the cylinder obstructs their passage to a singly-wound
circular loop. The existence of  such a motif might be relevant to the
constriction of membrane necks, indicating as it does that topology may
facilitate the constriction process.
\vskip1pc \noindent
The axial torque also depends in a surprisingly non-trivial way on the length of the loop.  It vanishes as this length become increasing large in all equilibrium loops. However, it does not do so monotonically: we discover the
existence of intriguing non-trivial states in which the axial torque vanishes
which occur when the length is tuned appropriately. These states admit an
analytic description. \vskip1pc \noindent The paper is organized as follows: In Sec. \ref{confinedelastica}, we develop the general framework for axially
symmetric binding. The first integral of the EL equation, associated with rotational symmetry, is derived in Sec. \ref{Sect:RotSym}. This framework is applied specifically to the cylinder in Sec. (\ref{SectCylConf}). The equilibrium states of a closed curve are constructed in Sec. \ref{Sect:Eqconfs}.  The axial torques, the energies, and  the forces they
transmit to the surface are determined. This is done perturbatively  for loop
states  approximating a multiply-covered circle and numerically otherwise. We
end with a discussion and a few suggestions for future work in Sec. \ref{Sect:Conclusions}. A number of useful identities and derivations are
collected in a set of appendixes where the analysis of vertical confined loops, loops with vanishing axial torque and force, and the
comparison of elastic curves on the cylinder with their planar counterparts are
also presented.

\section{Curves constrained to surfaces} \label{confinedelastica}

Consider a space curve in Euclidean space $\mathbb{E}^3$, $\Gamma: s
\rightarrow \mathbf{Y}(s)$ parametrized by arc-length constrained to lie on a
surface $\Sigma$. This surface is described in parametric form by the mapping
$\Sigma: (u^1,u^2) \rightarrow \mathbf{X}(u^1,u^2) \,\in \mathbb{E}^3$. The
tangent vectors adapted to this parametrization are ${\bf e}_a=\partial {\bf
X}/ \partial u^a$, $a=1,2$, the unit vector normal to the surface is ${\bf n}$,
the induced surface metric is $g_{ab}= {\bf e}_a\cdot {\bf e}_b$ and the
extrinsic curvature tensor is $K_{ab}= {\bf e}_a\cdot \partial_b {\bf n}$. The
confined curve can then also be described as a surface curve $\Gamma_\Sigma:
s\rightarrow (U^1(s),U^2(s))$. The surface-bound curve also carries a Darboux
frame, $\{{\bf T},{\bf l},{\bf n}\}$, where $\mathbf{T}= \mathbf{Y}'$ is the
unit tangent vector to $\Gamma$,\footnote{Here, and elsewhere, a prime represents
derivation with respect to the arc-length $s$.} $\mathbf{n}(U(s))$ is the
restriction of the unit
normal vector along the curve, and ${\bf l}= {\bf n} \times {\bf T}$ is the
conormal tangent to the surface, that is, the vector normal to the curve
$\Gamma$ but tangent to the surface. The structure equations--analogous  to the
Frenet-Serret (FS) equations--describing how the Darboux frame rotates along
the curve, are given by \cite{Kreyszig1991}
\begin{equation} \label{Darbouxequations}
\mathbf{T}' = \kappa_g \mathbf{l} - \kappa_n \mathbf{n}\,, \quad \mathbf{l}' =
-\kappa_g {\bf T} + \tau_g \mathbf {n}\,, \quad \mathbf{n}' = \kappa_n
\mathbf{T} - \tau_g {\bf l} \,,
\end{equation}
where the geodesic and normal curvatures, $\kappa_g$ and $\kappa_n$, as well as
the geodesic torsion $\tau_g$, are defined by
\begin{equation}
\kappa_g= {\bf T}' \cdot {\bf l}=  l^a t^b \nabla_b t_a \,,\qquad \kappa_n =
-{\bf T}'\cdot {\bf n} = t^a t^b \,K_{ab} \,, \qquad \tau_g = {\bf l}'\cdot{\bf
n} = -t^a l^b \, K_{ab} \,.
\end{equation}
Here $t^a$ and $l^a$ are the components of the vectors ${\bf T}$ and ${\bf l}$
with respect to the surface tangent basis ${\bf e}_a$, $a=1,2$: ${\bf T}= t^a
{\bf e}_a$, ${\bf l}= l^a {\bf e}_a$. $\nabla_a$ is the covariant derivative
compatible with $g_{ab}$. Whereas $\kappa_n$ and $\tau_g$ depend on the
extrinsic curvature, $\kappa_g$ is defined intrinsically; it depends only on
the surface metric $g_{ab}$. The FS frame is related to its Darboux counterpart through a rotation by an angle $\omega$ about the tangent direction. This
connection between the two frames provides a decomposition of the FS curvatures into its intrinsic and extrinsic parts
\begin{equation}\label{kaptau}
 \kappa_g = \kappa \, \cos \omega \,, \quad \kappa_n = \kappa \, \sin \omega
\,, \quad \tau = \tau_g - \omega'\,,
\end{equation}
so that $\kappa^2= \kappa_g^2 +\kappa_n^2$. The torsion $\tau$ is given by the
difference between the geodesic torsion and $\omega'$, the rotation rate of one frame with respect to the other. Note that $\tau_g$ involves two derivatives,
whereas $\tau$ involves three. The extra derivative is associated with the
derivative of $\omega$.
\vskip1pc \noindent
In equilibrium, one finds that the tension along the curves is balanced by a
normal force due to the confinement by the surface \cite{GuvVaz2012,
GuvValVaz2014},
\begin{equation} \label{EL}
\mathbf{F}' = - \lambda \mathbf{n}\,,
\end{equation}
where the tension in the loop ${\bf F}$ is  given by
\begin{equation} \label{FDarboux}
\mathbf{F} = \left(\frac{\kappa_g^2+\kappa_n^2}{2} - C \right) \mathbf{T} +
\left(\kappa_g' + \kappa_n \tau_g\right) \mathbf{l} - \left(\kappa_n' -
\kappa_g \tau_g \right) \mathbf{n}\,.
\end{equation}
${\bf F}$ represents the tension on a curve segment exerted by the segment with lower value of arclength \cite{CapoChryssGuv2002}; for instance, for a spiral
traversed upwards, it provides the force exerted on a segment by the region
below. The constant $C$ is associated with the constraint of fixed length and
can be identified with the Hamiltonian \cite{GuvValVaz2014}.
\vskip1pc \noindent
The tangential EL derivative $\varepsilon_{\bf T} = \mathbf{F}' \cdot \mathbf{T}$ vanishes identically, a consequence of the fact that the only
relevant degrees of freedom are geometrical.
\vskip1pc \noindent
The projection of Eq. (\ref{EL}) onto $\mathbf{l}$ provides the EL equation
\begin{equation} \label{ELeq}
\varepsilon_\mathbf{l} = {\bf F}' \cdot {\bf l} = \kappa_g'' + \kappa_g \left(
\frac{ \kappa_g^2 + \kappa_n^2}{2} - \tau_g^2 - C \right) +
\frac{\left(\kappa^2_n \tau_g\right)'}{\kappa_n} = 0\,.
\end{equation}
This equation was first derived in Ref. \cite{Nickerson1988} using a
procedure different from that in \cite{GuvVaz2012}. Note that it involves the
two curvatures as well as the geodesic torsion. In general, the curve will not
follow a geodesic with $\kappa_g =0$.
\vskip1pc \noindent
The corresponding projection onto ${\bf n}$ determines the magnitude of the
force $\lambda$ transmitted to the surface,
\begin{equation} \label{lambda}
\lambda = -{\bf F}' \cdot {\bf n} =\kappa_n'' + \kappa_n \left(
\frac{\kappa_g^2 + \kappa_n^2}{2} - \tau_g^2 -C\right) - \frac{\left(\kappa_g^2
\tau_g \right)'}{\kappa_g}\,.
\end{equation}
Thus the normal force is completely determined  when the local geometry is
known. Its magnitude will vary along the contact region even for confinement by a sphere. For a curve confined inside the surface, when $\lambda$ is positive
(negative) the curve is pushing (pulling) the surface. This expression is
missing in the framework presented in \cite{Nickerson1988}.  It was presented
in \cite{GuvVaz2012}.

\subsection{Axial symmetry and residual conserved torques} \label{Sect:RotSym}

The integrability of the EL equations for the unconstrained curve is a
consequence of the Euclidean invariance of its energy, which implies the
conservation of both forces and torques. The constrained counterpart will not
be integrable, for there will not be a sufficient number of conserved
quantities; in general, under confinement, one surrenders not only
translational invariance, but also rotational invariance. The torque about the
origin per unit length of the curve, ${\bf M}$, is  given by two contributions: the torque due to the force ${\bf F}$ about the curve's position and an
intrinsic torque,
\begin{equation} \label{eq:MFS}
{\bf M}= \ {\bf Y}\times {\bf F} + {\bf S}\,, \quad \mbox{where} \quad {\bf S}
= \kappa_g \, {\bf n} - \kappa_n \, {\bf l}\,.
\end{equation}
Like the tension vector ${\bf F}$, the vector ${\bf M}$ provides the torque on
a curve segment exerted by the one with a lower value of $s$. For a free  curve the torque vector is conserved, ${\bf M}'=0$, which can be cast in the
manifestly translational invariant form $\mathbf{T}\times \mathbf{F} +
\mathbf{S}' = 0$ \cite{CapoChryssGuv2002}. By contrast, for a confined curve
one has instead
\begin{equation} \label{Mp}
\mathbf{M}' = \varepsilon_\mathbf{l} \left(\mathbf{Y}\times \mathbf{l} \right)
- \lambda \left(\mathbf{Y} \times\mathbf{n}\right) \,.
\end{equation}
Thus, in a confined equilibrium, ${\bf M}$ will not generally be conserved,
even though $\varepsilon_\mathbf{l}=0$, because, in general, neither the normal force $\lambda$ vanishes, nor are ${\bf Y}$ and ${\bf n}$ parallel.
\vskip1pc \noindent
Let us now consider a curve confined by a surface with axial symmetry (about
the $Z$ axis, say). The axisymmetric surface is described by the radial and height coordinates $R$ and $Z$ of the generators, parameterized by the
arclength $\ell$.\footnote{Derivatives with respect to $\ell$ are denoted
by an overdot, $\dot{} = \partial_{\ell}$.} The azimuthal angle is denoted
by $\varphi$ (see Appendix \ref{Appaxisymsurf}). The curve is described by the
embedding $s\to (\ell(s),\varphi(s))$, or $\Gamma:s \rightarrow \mathbf {Y}(s)
= R(s) \hat{\bm \rho}(s) + Z(s)\hat{\bf z}$,  where $\hat{\bm \rho}(s) =
(\cos\varphi(s), \sin\varphi(s),0)$. The tangent and conormal vectors are given by
\begin{equation} \label{eqsTLAxs}
\mathbf{T} = \cos \alpha \, \hat{\bm \varphi}  + \sin \alpha \,
\mathbf{e}_{\ell}  \,, \qquad \mathbf{l} = -  \sin \alpha \, \hat{\bm \varphi}
+ \cos \alpha \, \mathbf{e}_{\ell}\,,
\end{equation}
where $\hat{\bm \varphi}(s) = (-\sin\varphi(s), \cos\varphi(s), 0)$ and ${\bf
e}_{\ell}(s) = \dot{R}(s) \, \hat{\bm \rho}(s) + \dot{Z}(s)\,\hat{\bf z}$;
$\alpha$ is the angle that the tangent of the curve makes with the azimuthal
direction $\hat{\bm \varphi}$. Comparison with the expression (\ref{Xlphiaxisym}) for the embedding functions of an axisymmetric surface gives
\begin{equation} \label{sincosaalpha}
\sin\alpha = \ell' \qquad \cos \alpha = R \, \varphi'\,.
\end{equation}
Using Eqs. (\ref{eqsTLAxs}) and (\ref{sincosaalpha}), the acceleration  ${\bf
T}'$ along the curve is given by
\begin{equation}
{\bf T}' = ({\alpha}' - \frac{R'}{R} \, \cot{\alpha}) \, {\bf l} - (\sin^2
\alpha \, \kappa_\perp + \cos^2 \alpha \, \kappa_\parallel) \, {\bf n}\,,
\end{equation}
where $\kappa_\perp$ and $\kappa_\parallel$ are the principal curvatures along
the meridians (generators with constant $\varphi$) and the parallels (circles
of constant $\ell$), respectively (see Appendix \ref{Appaxisymsurf}). Thus the  geodesic
curvature of $\Gamma$, $\kappa_g = {\bf T}' \cdot {\bf l} $, is given by
\begin{equation} \label{eq:kappagaxsym}
\kappa_g = - \frac{(R\,\cos \alpha)'}{R\,\sin \alpha}\,;
\end{equation}
the corresponding normal curvature  $\kappa_n = -{\bf T}' \cdot {\bf n}$  is
given by Euler's equation for the axisymmetric surface,
\begin{equation} \label{kappanaxcurvprin}
\kappa_n = \sin^2 \alpha \, \kappa_\perp + \cos^2 \alpha \,\kappa_\parallel\,.
\end{equation}
Likewise the geodesic torsion $\tau_g$ is given by
\begin{equation} \label{eq:taugaxsym}
 \tau_g = \sin \alpha \cos \alpha (\kappa_\parallel - \kappa_\perp)\,.
\end{equation}
Using the above expressions for the Darboux frame of a curve on an axisymmetric surface, along with the expression $\dot{Z} = R \, \kappa_\parallel$ (see
Appendix \ref{Appaxisymsurf}) and the identity $\cos \alpha \, \kappa_\parallel = - \cos \alpha \, \kappa_n - \sin \alpha \, \tau_g$, obtained by a linear
combination of Eqs. (\ref{kappanaxcurvprin}) and (\ref{eq:taugaxsym}), one
finds that the torque about the symmetry axis, $M^Z = \mathbf{M}\cdot \hat{\bf
z}$, is given by \cite{GuvValVaz2014}
\begin{equation} \label{M3eq}
M^Z = R \left( -\sin \alpha \, (\kappa_g' + 2 \,\kappa_n \, \tau_g) + \cos
\alpha \, \left(\frac{\kappa_g^2- \kappa_n^2}{2} - C\right) \right) + \dot{R}
\,\kappa_g \,.
\end{equation}
It is easy to confirm that this component is conserved: projecting Eq.
(\ref{Mp}) onto $\hat{\bf z}$, one gets
\begin{equation}
 M^{Z\,'} = {\bf M}' \cdot \hat{\bf z} = {\bf Y} \cdot \left( \varepsilon_{\bf
l} {\bf l} \times \hat{\bf z} - \lambda {\bf n} \times \hat{\bf z}\right)\,.
\end{equation}
In an axisymmetric surface the three vectors ${\bf Y}$, ${\bf n}$ and $\hat{\bf z}$ are coplanar and ${\bf l} \times \hat{\bf z} = -\sin \alpha \hat{\bm \rho}- \cot \alpha R' \hat{\bm \varphi}$; therefore, $M^{Z\,'} = -\sin \alpha \, R \,
\varepsilon_{\bf l} $, so it vanishes in equilibrium as claimed. Equation
(\ref{M3eq}) provides a first integral of the EL equation (\ref{ELeq}), which
is not obvious by inspection.
\vskip1pc \noindent
To reconstruct the curve, the second order differential equation (\ref{M3eq})
needs to be solved for the angle $\alpha$ with fixed $M^Z$ and $C$,
and appropriate boundary conditions (BCs), which could be specifying the
initial angle and its derivative (Cauchy BCs) or the angle at the boundaries
(Dirichlet BCs). Once $\alpha$ is known, the position on the surface is
determined using the relations (\ref{sincosaalpha}). The two constants $M^Z$
and $C$ can be tuned to fix the length of the curve, $L$, and total azimuthal
angle turned, $\Delta \varphi$.

\section{Cylindrical Confinement} \label{SectCylConf}

Now we apply this framework to treat curves lying on a cylinder. Despite the
apparent simplicity of the cylindrical geometry, the determination of the
equilibrium configurations of curves on a cylinder is not straightforward.
However, as explained below, the additional translational symmetry facilitates
the integration of the EL equation.

\subsection{Derivation of the quadrature}

The height function is given by the meridian arc length, $Z=\ell$. We normalize all lengths in terms of the cylinder radius $R_0$. Scaled quantities are
denoted by lowercase letters or by an overbar; e.g., the scaled height function is $z := Z/R_0$. Curves on the cylinder are parameterized by scaled arc
length $s/R_0$.\footnote{Since there is no possible confusion, the scaled
arc-length is also denoted by $s$.}
\vskip1pc \noindent
The scaled geodesic curvature is given by
\begin{equation} \label{kgcylinder}
\bar{\kappa}_g := R_0 \, \kappa_g = \alpha'\,,
\end{equation}
The two scaled principal curvatures are $\bar{\kappa}_\perp := R_0 \,
\kappa_\perp = 0$ and $\bar{\kappa}_\parallel := R_0 \, \kappa_\parallel = 1$,
so that the scaled normal curvature and geodesic torsion are given by
\begin{equation} \label{kncylinder}
\bar{\kappa}_n := R_0 \, \kappa_n = \cos^2 \alpha\,; \quad \bar{\tau}_g := R_0
\, \tau_g = \sin \, \alpha \cos \alpha\,.
\end{equation}
Inserting these expressions into the EL equation (\ref{ELeq}), it
is easily seen to read ($c = R_0^2 \, C$)
\begin{equation} \label{ELcylcase}
\varepsilon_{\bf l} = \alpha''' + \alpha' \left(\frac{\alpha'{}^2}{2} +
\frac{3}{2}\cos^4 \alpha - 6 \,\sin^2  \alpha \cos^2 \alpha - c \right)  =0\,.
\end{equation}
The scaled first integral, Eq. (\ref{M3eq}) (${\bf m} := R_0 \, {\bf M}$), reads
\begin{equation}\label{M3cyl}
m := {\bf m} \cdot \hat{\bf z} = - \sin \alpha \left(\bar{\kappa}_g' + 2
\bar{\kappa}_n \, \bar{\tau}_g \right) + \cos \alpha
\left(\frac{\bar{\kappa}_g^2 - \bar{\kappa}_n^2}{2} - c\right) \,.
\end{equation}
One could now solve this third order differential equation (second order for
$\alpha$) following the procedure outlined at the end of the previous section.
However, it is possible to do better. We have yet to  exploit the translational invariance along the cylindrical axis. When we do, we obtain a quadrature for
$\alpha$.
\vskip1pc \noindent
Translational invariance along $\hat{\bf z}$ implies that the projection of the scaled stress vector ${\bf f} := R^2_0 {\bf F}$ along the cylinder axis is
constant. Projecting ${\bf f}$ onto $\hat{\bf z}$, and making use of
definitions of ${\bf T}$ and ${\bf l}$ given in Eq. (\ref{eqsTLAxs}), one
identifies the first integral,
\begin{equation} \label{F3cyl}
 f := {\bf f} \cdot \hat{\bf z} = \cos \alpha \left(\bar{\kappa}_g' + 2
\bar{\kappa}_n \bar{\tau}_g \right) + \sin \alpha \left(\frac{\bar{\kappa}_g^2
- \bar{\kappa}_n^2}{2} - c \right)\,.
\end{equation}
It is straightforward to confirm that $f' =  R^2_0 \, \cos \alpha \,
\varepsilon_{\bf l}$.
\vskip1pc\noindent
Both first integrals are of second order in derivatives of $\alpha$ (this
dependence enters through $\bar{\kappa}_g'$ in Eqs. (\ref{M3cyl}) and
(\ref{F3cyl})). One can now take an appropriate linear combination to eliminate this second derivative: specifically, the definition
 \begin{equation} \label{2ndintcylconf}
 \mu (\alpha) := m \,  \cos \alpha + f \, \sin \alpha  = \frac{\bar{\kappa}_g^2
- \bar{\kappa}_n^2}{2} - c \,,
\end{equation}
will lead to a quadrature for $\alpha$. Substituting for $\kappa_g$ and
$\kappa_n$ one gets:
\begin{equation} \label{CylQuad}
 \frac{1}{2} (\alpha')^2 + U(\alpha) = c\,, \quad \mbox{where} \quad U(\alpha)
= - \frac{\cos^4 \alpha}{2} -  \mu(\alpha)\,.
\end{equation}
It is not obvious at the level of the EL equation that two integrations are possible. However, it should be remarked that the poor man's derivation of the quadrature using a variational principle adapted to symmetry demystifies this ``coincidence''. It is provided in  Appendix \ref{Appcylhamform}, a special instance of a more general construction presented in \cite{GuvValVaz2014}.
\vskip1pc \noindent
One may ask if any additional information is to be gleaned by taking some other linear combination of $f$ and $m$. Consider
\begin{equation} \label{Cylfrstint}
\nu (\alpha) := -m \, \sin \alpha + f \, \cos \alpha = \bar{\kappa}_g' + 2 \,
\bar{\kappa}_n \, \bar{\tau}_g\,.
\end{equation}
In terms of $\alpha$ this second order differential equation reads
\begin{equation} \label{1stintcylconf}
 \alpha'' + 2\cos^3 \alpha \sin \alpha - \nu(\alpha) = 0\,.
\end{equation}
On the one hand, taking into account expression (\ref{kgcylinder}), one has
$\mu' = \nu \kappa_g$. On the other hand, using the identity for the cylinder
$\bar{\kappa}'_n = -2 \bar{\kappa}_g \, \bar{\tau}_g$, the derivative of the
right-hand side of Eq. (\ref{2ndintcylconf}) can be written as
\begin{equation} \label{kappagnsqrdid}
 \left(\frac{\bar{\kappa}_g^2 - \bar{\kappa}_n^2}{2} - c \right)' =
(\bar{\kappa}_g' + 2 \, \bar{\kappa}_n \, \bar{\tau}_g) \kappa_g\,.
\end{equation}
These two relations confirm that Eq. (\ref{Cylfrstint}) follows from
differentiation of Eq. (\ref{2ndintcylconf}). Alternatively, substituting
identity (\ref{kappagnsqrdid}) in Eqs. (\ref{M3cyl}) and (\ref{F3cyl})) for $m$ and $f$, and integrating them, one obtains the quadrature (\ref{2ndintcylconf}). Further differentiation of Eq. (\ref{Cylfrstint}) and substitution of the EL equation to replace $\kappa''_g$ in favor of terms quadratic in the Darboux curvatures only reproduces Eq. (\ref{2ndintcylconf}).
\vskip1pc \noindent
The quadrature (\ref{CylQuad}) provides us with an analogue of a particle with
unit mass and energy $c$ in a potential $U(\alpha)$.  While the hard work lies
ahead, this will prove to be very useful throughout this paper. That $c$ plays
the role of the energy in this identification is particularly appropriate,
because $c$ is also the conserved quantity, i.e., the Hamiltonian function,
associated with the absence of an explicit dependence on $s$ in the bending
energy density, as detailed in Appendix \ref{Appcylhamform}.
\vskip1pc \noindent
The first term in $U$, proportional to $\cos^4 \alpha$ (or $\bar{\kappa}^2_n$),
reflects the anisotropy of the sectional curvatures on the cylinder, which
breaks the rotational symmetry about the normal vector. As a result, despite
the fact that the cylinder is isometric to a plane, its extrinsic geometry is anisotropic and elastic curves on a cylinder behave differently from their
counterparts on the plane. It is precisely this term that distinguishes
elastic curves on the cylinder from their planar Euler elastic counterparts,
which are integrable in terms of elliptic functions, as described in Appendix
\ref{SectConfPEE}. This can be seen by introducing the change of variables
$\alpha \rightarrow \psi + \beta$, $f \rightarrow F \sin \beta$ and $m
\rightarrow F \cos \beta$. Now one has $\mu \rightarrow F \, \cos \psi$ and
$\nu \rightarrow - F \sin \psi$, so without the $\kappa^2_n$ term and its
derivative, the quadrature (\ref{CylQuad}) and the first integral
(\ref{Cylfrstint}) reduce to Eqs. (\ref{factquad}). If the curve lies close to
a parallel or the cylinder has a very large radius, $R_0 \gg 1$, so that
$\kappa_n \rightarrow 0$, $\tau_g \rightarrow 0$ and $\kappa_g$ becomes the FS
curvature of the curve on the plane, the two descriptions coincide. The
comparison between the two is discussed in Sec. \ref{Sect:nonlin}.
\vskip1pc \noindent
The two coordinate functions of the curve on the cylinder, are determined by
the relations (\ref{sincosaalpha}), which in terms of the scaled quantities
reduce to
\begin{equation} \label{sincosaalphacyl}
 z' = \sin \alpha \,, \quad \varphi' = \cos \alpha\,.
\end{equation}

\subsection{Total bending energy and constraining force}

The scaled bending energy of the constrained curve is given  in terms of the
angle $\alpha$  by
\begin{equation}
h_B := R_0 \, H_B = \frac{1}{2} \int {\rm d}s \,( \alpha' {}^2 + \cos^4
\alpha)\,.
\end{equation}
Using the quadrature (\ref{2ndintcylconf}), the derivative can be eliminated
in favor of $\alpha$, so that
\begin{equation}
h_B = \int {\rm d}s \,\left( \cos^4 \alpha + f \sin \alpha + m \cos \alpha
+c\right) \,.
\end{equation}
Taking into account the relations $z' = \sin \alpha$ and $\varphi' = \cos
\alpha$, the last terms can be integrated
\begin{equation} \label{hBDeltazphiL}
h_B = \int {\rm d}s \, \cos^4 \alpha + f \Delta z + m \Delta \varphi + c \, l
\,,
\end{equation}
where $\Delta z=\Delta Z/R_0$, $\Delta \varphi$ and $l=L/R_0$ are the total
scaled height, winding and scaled length of the curve. For a closed curve
$\Delta z=0$ and $\Delta \varphi = 2 \pi p$. This relation makes explicit the
respective interpretations of $f$, $m$, and $c$ as coupling to the height,
rotational extent and length, respectively.
\vskip1pc \noindent
Likewise, using the quadrature (\ref{CylQuad}) and Eq. (\ref{1stintcylconf}) to eliminate derivatives of $\alpha$ in the expression (\ref{lambda}) for the
magnitude of the scaled local constraining force on the cylinder,
$\bar{\lambda} = R_0^3 \lambda$, can be expressed as
\begin{eqnarray} \label{eq:Fcyl1munu}
\bar{\lambda} &=& 2 \, \bar{\kappa}_n^2  \left(5 - 6 \, \bar{\kappa}_n\right)+
\left(6 -11 \, \bar{\kappa}_n\right) \mu -4\, \bar{\tau}_g^2 \, \nu + 6\,c\,
\left( 1 - 2 \, \bar{\kappa}_n\right) \,,
\end{eqnarray}
or in terms of $\alpha$ and constants $m$, $f$ and $c$,
\begin{equation} \label{eq:Fcyl1mf}
\bar{\lambda} = 2 \, \cos^4 \alpha \left(5-6 \cos^2 \alpha \right) - 6\,c\,
\cos \,2 \alpha - 3\,f \sin \alpha \left(5 \cos^2\alpha - 2\right) - 5\,m
\cos\,\alpha \left(3 \cos^2 \alpha -2\right) \,.
\end{equation}
It depends only on the local value of the turning angle.

\section{Closed equilibrium configurations} \label{Sect:Eqconfs}

Now we focus on isolated loops, i.e., closed curves without external
sources of tension other than those due to confinement.\footnote{Recall that we ignore polymer twist, so one might wonder whether this could change the solutions. However, if the polymer is rotationally symmetric, having no ``preferred side'' for it to adhere onto the surface and no preferred bending direction, any local twist does not couple to the surface constraint, and so we only need to consider the \emph{global} constraint. If we now start with an untwisted circle and imagine extending its length in such a way that no twist is added, ignoring twist right from the start is permissible.} Let the loop have length $L=2 \pi R$, $p$ be the number of times it winds the cylinder and $n$ the number of periods in completing a circuit of the loop. Loops are characterized by $p$ and $n$: loops with different $p$ are homotopically inequivalent; $n$ distinguishes equilibrium states within a given homotopy class. We denote the sequence of loops with fixed values of these two integers, generated by varying the length of the loop by ${\rm L}_{p,n}$. Define the scaled total length $l = L/R_0 = 2 \pi r$, where $r=R/ R_0$ is the scaled radius of the loop. Since $l\geq 2 \pi p$, the scaled radius is bounded from below by the number of windings, $r \geq p$. The length is expressed in terms of the scaled excess radius $\Delta r = r - p \geq 0$.
\vskip1pc \noindent
We examine loops that oscillate symmetrically about a parallel, which we
set at $z=0$ and refer to as the \emph{equator}. This implies that the odd term
linear in $\sin\alpha$ vanishes, so $f=0$.
Note that  the potential $U$ appearing in (\ref{CylQuad}) now possesses up-down
symmetry, $\alpha \to -\alpha$. However, the constant $m$ does not necessarily
vanish  which breaks the left-right symmetry $\alpha \to \pi - \alpha$.
\vskip1pc \noindent
The tangent angle $\alpha$ will be bounded in a closed loop. It will now
oscillate symmetrically between the turning points of the even potential, $-
\alpha_M$ and $\alpha_M$. These turning points will occur at the intersection
of the loop with the equator;  $\alpha=0$ occurs on the two extremal parallels.
\vskip1pc \noindent
The two parameters $m$ and $c$ are determined by imposing the periodicity on
the coordinates, required by the closure constraint. Note, first of all, that
closure in the axial direction is consistent with $f=0$. To see this, note that
for a closed curve, after one half period one has $\Delta z = 0$
\cite{Langer1984, SingerSantiago2008};  using the quadrature (\ref{CylQuad})
and the relation (\ref{sincosaalphacyl} connecting $z$ to $\alpha$  one has
\begin{equation}
  \Delta z = \frac{1}{\sqrt{2}} \, \int_{-\alpha_M}^{\alpha_M} {\rm d}\alpha \,
\frac{\sin \alpha}{\sqrt{c - U(\alpha)}} = 0\,. \label{Deltaz}
\end{equation}
This is because $U(\alpha)$ is an even function of $\alpha$ so that the
integrand is odd.
\vskip1pc \noindent
The remaining two parameters $m$ and $c$ can be related to the three geometric
quantities $p$, $n$ and $\Delta r$ by integrating the quadrature
(\ref{CylQuad}) along with the relation connecting $\varphi$ to $\alpha$, (Eq.
(\ref{sincosaalphacyl}))
using the fact that the azimuthal range $\Delta \varphi$ is quantized; i.e.,
$\Delta \varphi/ (2\pi)=p$.  One obtains
\begin{subequations} \label{Deltarphi}
\begin{eqnarray}
\Delta r &=& \frac{4 \,n}{\sqrt{2} \,2 \, \pi} \int_{0}^{\alpha_M}\, {\rm
d}\alpha \, \frac{1}{\sqrt{c-U(\alpha)}} -p \,, \label{Deltar}\\
p &=& \frac{4\,n}{\sqrt{2} \, 2 \,\pi } \, \int_{0}^{\alpha_M} \, {\rm d}\alpha
\, \frac{ \cos \alpha}{\sqrt{c - U(\alpha)}}   \,.\label{Deltaphi}
 \end{eqnarray}
\end{subequations}
In Eq. (\ref{Deltaphi}) we have used the fact that $\alpha$ completes $n$
periods while the loop completes $p$ trips around the cylinder.
\vskip1pc \noindent
Before analyzing the loops in the nonlinear regime, we first solve the
quadrature in a perturbative manner for loops which deviate slightly from a
circular loop covering  the equator $p$ times. The insight we obtain will prove useful when solving the quadrature (\ref{CylQuad}) numerically.

\subsection{Perturbative solutions about a circular loop} \label{sectpertsols}

To examine deformations of a circular loop perturbatively, we let $\alpha_1$
represent a small perturbation about $\alpha_0=0$. We set $f=0$ and expand the
constants, $m = m_0 + m_2 + \dots$ and $c = c_0 +  c_2  +\dots$,\footnote{First order terms are not considered because they will be proportional to first
order changes in the length of the circle, which  vanishes on account of the
geodesic character of the parallel.} . The subindices represent the order in
the expansion. The approximation of the quadrature (\ref{CylQuad}) at lowest
order reads
\begin{equation} \label{m01}
m_0 = -c_0 -\frac{1}{2} \,.
\end{equation}
Note that $m_0$ and $c_0$ appear in the combination $m_0+c_0$ which is constant independent of $n$ and $p$. At second order the quadrature describes harmonic oscillations,
\begin{equation} \label{harmcylquad}
\frac{1}{2} \left(\alpha'_1\right)^2 + V_2(\alpha_1)  = E_2 \,,
\end{equation}
where we have defined
\begin{equation} \label{V2}
V_2(\alpha_1) = a_0 \, \alpha^2_1\,, \quad  a_0 =  m_0/2 + 1\, \quad \mbox{and}
\quad E_2 = m_2 + c_2\,.
\end{equation}
Thus, the behavior of the perturbation can be described in analogy to a particle of unit mass and energy $E_2$ moving in a quadratic potential $V_2$.  The parameter $a_0$ is necessarily positive. Otherwise, $V_2$ would be a downward parabola, $\alpha$ would not be periodic, and the corresponding loops would be open.
\vskip1pc \noindent
The two turning points of $V_2$, corresponding to the maximum and minimum
values of $\alpha$, are $\pm \alpha_{M\,1}$, where
\begin{equation} \label{amE2qa0}
\alpha_{M\,1} = \frac{\sqrt{2 E_2 }}{q} \, \quad \mbox{and} \quad q^2 := 2 a_0
= m_0 + 2 \,.
\end{equation}
Solving Eq. (\ref{harmcylquad}) one identifies
\begin{equation}
 \alpha_1 = \alpha_{M\,1} \, \cos q \, s\,;
\end{equation}
combining Eqs. (\ref{m01}) and (\ref{amE2qa0}), $m_0$ and $c_0$ can be
expressed in terms of $q$
\begin{equation}  \label{mc0pert}
m_0 = q^2 - 2 \,, \quad c_0 = \frac{3}{2} - q^2 \,.
\end{equation}
The condition that the periodic perturbation should complete $n$ periods, while the curve winds $p$ times around the cylinder implies that the wave number is
given by the ratio of periods to windings,
\begin{equation}
q = \frac{n}{p}\,.
\end{equation}
Notice that $m_0$ does not vanish because $q$ is a rational number. A critical
axial torque is required to deform a loop of radius $R_0$ into a state with a
given ratio $q$.  In particular, $m_0$ is negative when $n=1$ and positive for
all higher values of $n$.  This sign change has physical significance.
Superficially, states with the same $p$ and different values of $n$ may appear
to be identical when $\Delta r=0$. However, as we will see, the different
values of $m_0$ and $c_0$ in these states determine the critical buckling
forces required to initiate the deformation of the circular loop into the
appropriate state.
\vskip1pc \noindent
From the relation (\ref{amE2qa0}) between $E_2$ and $\alpha_{M\,1}$, one
concludes that the sum of their second order corrections is constrained by the
amplitude
\begin{equation} \label{E2eq}
E_2 = m_2 + c_2 = \frac{q^2}{2} \, \alpha^2_{M\,1}\,.
\end{equation}
To identify a second relation between $m_2$ and $c_2$, we  use the conditions
(\ref{Deltarphi}),  which relate the excess radius and the azimuthal range in
terms of the potential. Since to first order the turning points are $\pm
\alpha_{M\,1}$, the denominator in these expressions can be expanded as
\begin{equation}
 c- U(\alpha) \approx (\alpha_{M\,1}^2 - \alpha_1^2) (\chi - \xi \alpha_1^2)\,,
\end{equation}
where
\begin{equation}
\chi = \frac{1}{2} \left(q^2 + m_2\right) - \xi \, \alpha_{M\,1}^2 \,, \quad
\xi = \frac{1}{24} \left(18 + q^2\right)  \,;
\end{equation}
thus,
\begin{equation}
 \frac{1}{\sqrt{2}\, \sqrt{c- U(\alpha)}} \approx \frac{1}{q
\sqrt{\alpha_{M\,1}^2 - \alpha_1^2}}  \, \left[1
+\frac{1}{q^2}\left(\xi\,\left( \alpha_{M\,1}^2 + \alpha_1^2 \right)-
\frac{m_2}{2}  \right)\right] \,.
\end{equation}
Substituting this expression into the two conditions (\ref{Deltarphi}) and
integrating, one obtains
\begin{equation}  \label{Dltrm2}
\Delta r = \frac{p}{2 \, q^2} \, \left(3 \, \xi \, \alpha_{M\,1}^2 -
m_2\right)\,, \quad m_2= \left(3 \, \xi - \frac{q^2}{2} \right) \,
\alpha^2_{M\,1}\,.
\end{equation}
From these two equations one finds that the excess radius is proportional to
the square of the perturbation amplitude,
\begin{equation} \label{DeltaralphaM}
\Delta r =  \frac{p}{4}\,\alpha_{M\,1}^2\,.
\end{equation}
Using this relation in Eqs. (\ref{E2eq}) and (\ref{Dltrm2}), one can express
$m_2$ and $c_2$ in terms of the excess radius:
\begin{equation}
 m_2 = \frac{3}{p}\, \left(3 - \frac{1}{2} \, q^2 \right) \Delta r\,, \quad c_2 = \frac{1}{p}\, \left(\frac{7}{2} \, q^2 - 9 \right) \Delta r\,.
\end{equation}
To complete the construction of the constrained curve, one needs to determine
its position coordinates $z$ and $\varphi$ by integrating the relations
(\ref{sincosaalphacyl}). To first order,
\begin{equation} \label{1srordcylzpphip}
z'_1  = \alpha_1 \,, \quad \varphi'_1 = 0\,;
\end{equation}
consequently, this confirms to first order that  a non-vanishing value of the conserved axial force $f$ is inconsistent with closed deformations of a parallel circle. Note first that, if $f\ne 0$, the potential $V_2$ defined in Eq. (\ref{V2}) gains a linear term so that $z_1$ picks up a helical contribution proportional to arclength, which is inconsistent with closure:
\begin{equation}
z_1 = \frac{2}{n} \, \sqrt{p\, \Delta r}  \, \sin q s \,, \quad \varphi_1 = 0
\,.
\end{equation}
In contrast to its spherically confined counterpart, the elliptic deformation
of an elastic loop on a cylinder with $n=1$ and $p=1$ is not identified with a
rotation, but with a tilt of the loop extending its length.

\subsubsection{Energy and confining force of perturbed loops}

In the quadratic approximation the bending energy $h_B$ is given by
\begin{equation}
h_B  \approx  \pi \left(p +  \left(2 \, q^2 -3 \right) \, \Delta r  \right)\,.
\end{equation}
Using the fact that $r^{-1} \approx p -\Delta r$, the energy can also be
expressed  in terms of the energy of the circle with reduced radius $r$, $h_c = \pi/ r $, as
\begin{equation} \label{hhc}
 h_B  \approx  h_c\, \left[ 1 + 2 \, \left( q^2  - 1 \right) \, \Delta
r\right]\,.
\end{equation}
It is now manifest that at lowest order the energy of the deformed loop is
given by the energy of the circular loop, $h_B \approx h_c$. The second order
correction to the energy of the loop vanishes for elliptic perturbations with
$q=1$, whereas it increases (decreases) if $q>1$ ($q<1$).
\vskip1pc \noindent
To second order, the magnitude of the normal force $\bar{\lambda}$ given by Eq.
(\ref{eq:Fcyl1mf})  reads
\begin{equation} \label{lambdapert}
 \bar{\lambda} \approx q^2 -1 + \frac{1}{p}\,\left[q^2 \left(11 \cos 2 q
\varphi - \frac{5}{2}\right) + 7 -2 \cos 2 q \varphi \right] \Delta r\,.
\end{equation}
For the elliptical confined states with $q=1$,  only the second order
correction is non-zero, $\bar{\lambda} \approx 18  \, \left( \cos^2 \varphi -
\frac{1}{4}\right) \, \Delta r$. It vanishes in the limit $\Delta r \to 0$.
Note that $\bar{\lambda}$ is not positive everywhere, assuming negative values
in the two intervals  $\varphi \in (\pi/3 , 2/3 \pi)$ and $\varphi \in (-2/3
\pi, - \pi/3)$,  describing the neighborhood of the turning points where the
height is  a maximum and a minimum. On account of this, if the loop is bound
inside the cylinder and free to detach, one expects that it will tend to rotate into the vertical plane, touching the cylinder tangentially at two points. Such states are described exactly by two segments of a planar Euler elastic curve as described in Appendix \ref{SectConfPEE}. A loop bound to the outer side of the
cylinder, free to detach, will tend to detach on the equator first before
reforming as a circular loop.
\vskip1pc \noindent
Note that a constant normal force $\bar{\lambda}_0 = q^2 -1$ persists in the limit $\Delta r \to 0$ in all other states. This behavior is analogous to the Euler instability associated with $n$-folds confined by a sphere
\cite{GuvVaz2012}.\footnote{Its relation to the non-vanishing  axial torque
($m_0$) associated with the confined loop in this limit is somewhat mysterious.}
In general, $\bar{\lambda}_0$ is positive (negative) if $q>1$ ($q<1$), so
initially the loop will push (constrict) everywhere on the cylinder. In particular, an interior bound curve with $p=1$ and $n$-fold dihedral symmetry will push on the cylinder, whereas its exterior counterpart will peel off.
\vskip1pc \noindent
Whereas the second order correction to the transmitted force can have either
sign (correlating with the sign of the cosine of the double angle), one has to
proceed beyond perturbation theory to determine if the sign of the force in
non-elliptical states changes in longer loops, as we will see in Sec.
\ref{Sect:nonlin}, it does.
\vskip1pc \noindent
An interior bound figure-eight ($n=1$, $p=2$) would be expected to unwind
into the planar Euler elastic vertical confined state; an exterior one, on the
other hand, will constrict the cylinder. Note that this configuration possesses a lower energy than a free twice-covered circular loop. There is no inclination to detach from the cylinder.  With increased length, as we see in our
non-perturbative treatment of the problem, we would not expect this state to
remain completely attached; intuitively  we would nonetheless expect it to
continue to constrict the cylinder. The topological obstruction provided by the cylinder prevents the doubly wound elastic loop from unraveling into its singly wound ground state. This will be confirmed when we venture beyond perturbation
theory. An exterior bound loop wound more than once around a cylinder will
always constrict the cylinder.

\subsubsection{Stability of perturbed loops} \label{sectstability}

\begin{figure}[htb]
\begin{center}
\includegraphics[scale=0.9]{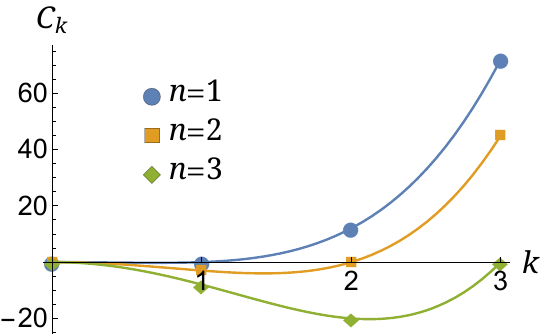}
\caption{(Color online) Eigenvalues $C_k$ corresponding to deformations of a parallel circle. For all $n>1$,  the spectrum possesses a negative eigenvalue.} \label{Fig2}
\end{center}
\end{figure}

Modulo the EL equation, the second variation of the energy is given by
\begin{equation}
 \delta^2 H_B = \int \, {\rm d}s \, \Phi {\cal L} \Phi \,,
\end{equation}
where ${\cal L}$ is a self-adjoint fourth order differential operator, defined
at lowest order and about a $p$-fold covering of the equator by
\begin{equation}
 {\cal L} = \frac{\partial^2}{\partial s^2} \left( \frac{\partial^2}{\partial
s^2} + q^2 \right)\,,
\end{equation}
and $\Phi$ is the normal deformation along $\mathbf{l}$,  $\Phi= \delta {\bf Y} \cdot  {\bf l}$. On account of the fixed length condition, the permissible normal deformations ought to be orthogonal to the geodesic curvature, so $\Phi$ must satisfy the global constraint \cite{GuvMulVaz2012}
\begin{equation} \label{kgPsiperort}
 \oint {\rm d}s \,  \kappa_g \, \Phi = 0 \,.
\end{equation}
The eigenmodes of ${\cal L}$ are $1, \varphi, \cos Q \varphi, \sin Q
\varphi$.\footnote{Recall that to lowest order $s$ is given by $\varphi$.}.
Since the periodic eigenmodes ought to have the same number of windings as the
original loop, the wave number is given by $Q=k/p$, where $k$ is another
integer. Their corresponding eigenvalues are
\begin{equation}
 C_k = \frac{k^2}{p^4} \left(k^2-n^2\right)\,.
\end{equation}
There are four zero modes with $C_k=0$, the constant and linear modes $1$ and
$\varphi$ and two (cosine and sine) eigenmodes with the same number of periods as the original loop, $k=n$ ($Q=q$). However, the eigenmode $\varphi$ is not
periodic and the eigenmode $\cos q \varphi \propto \kappa_g$ does not satisfy
the global isometry condition (\ref{kgPsiperort}), so both of these modes are
not considered. The allowed zero modes correspond to a translation along
($\Phi_T=1$) and a rotation about ($\Phi_R = \sin q \varphi \propto \kappa_g'$)
the axis of the cylinder.
\vskip1pc \noindent
In Fig. \ref{Fig2} the eigenvalues $C_k$ are plotted for the first $n$-fold
perturbations of a single covering of the equator ($q=1$). These eigenvalues
are all positive for the elliptic state with $n=1$. All states with $n\ge 2$,
however, exhibit negative eigenvalues, which signal the possibility of decay to states with lower energy, if they are accessible. Therefore, in this
approximation only the elliptical state $n=1$ is stable, whereas the higher
$n$-folds with $n\ge 2$ are unstable.  Of course, to access the stability of
the elliptic ground state one need to also treat perturbations lifting the loop off the surface. In the absence of an adhesive force one would expect this
state   to be unstable with respect to such perturbations.

\subsection{Non-linear regime} \label{Sect:nonlin}

\begin{figure}[htb]
\begin{center}
\includegraphics[scale=0.8]{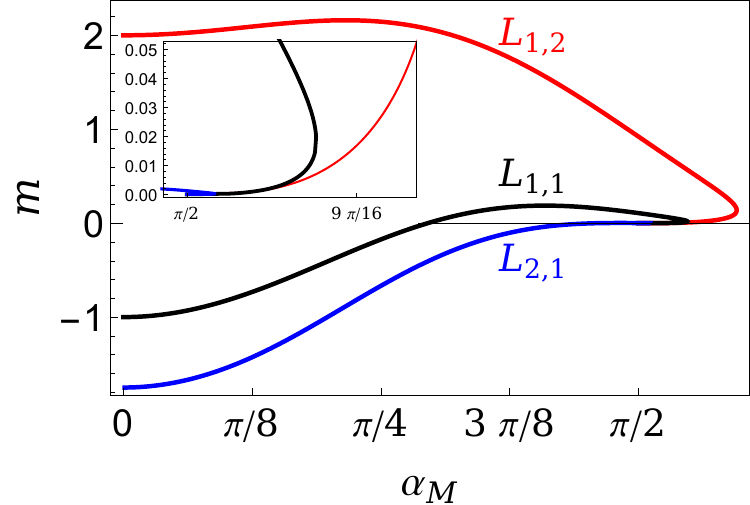}
\end{center}
\caption{(Color online) Behavior of the vertical torque $m$ as a function of
$\alpha_M$ for the sequences ${\rm L}_{2,1}$ (blue curve), ${\rm L}_{1,1}$ (black curve)
and ${\rm L}_{1,2}$ (red curve). The complicated relationship between $m$ and
$\alpha_M$ upon the development of vertical tangents  with $\alpha_M\pi/2$
(detailed in the inset) is clarified by examining separately the functional
dependencies of $m$ and $\alpha_M$ on $\Delta r$.} \label{Fig3}
\end{figure}

\begin{figure}[htb]
\begin{center}
  \subfigure[]{\includegraphics[scale=0.72]{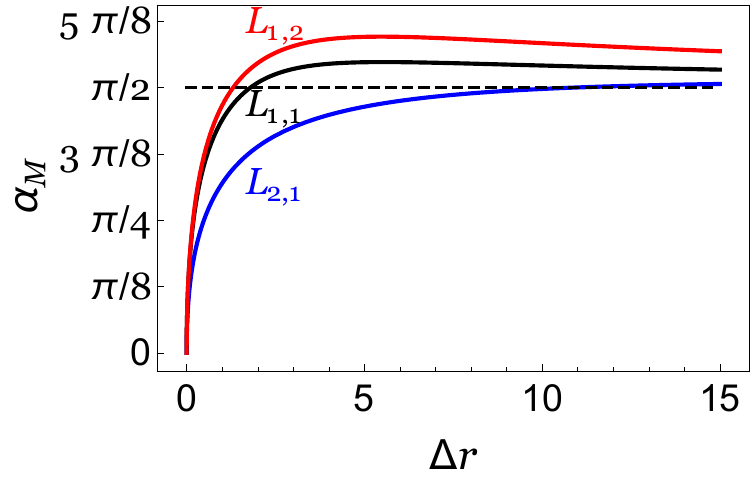}}
\subfigure[]{\includegraphics[scale=0.68]{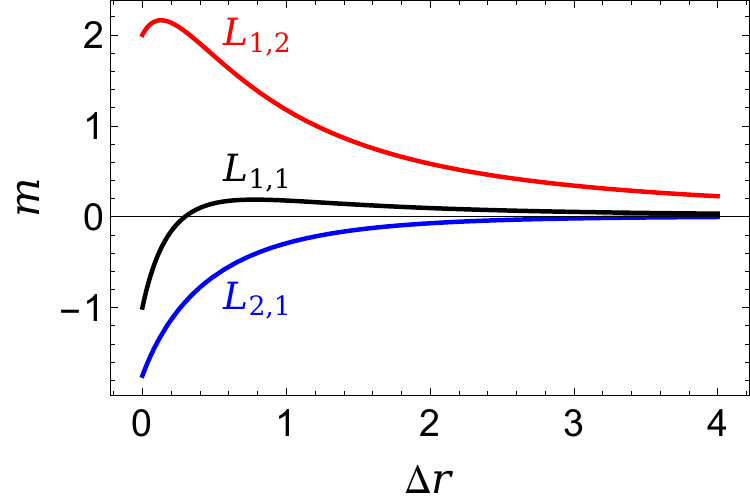}}\\
  \subfigure[]{\includegraphics[scale=0.68]{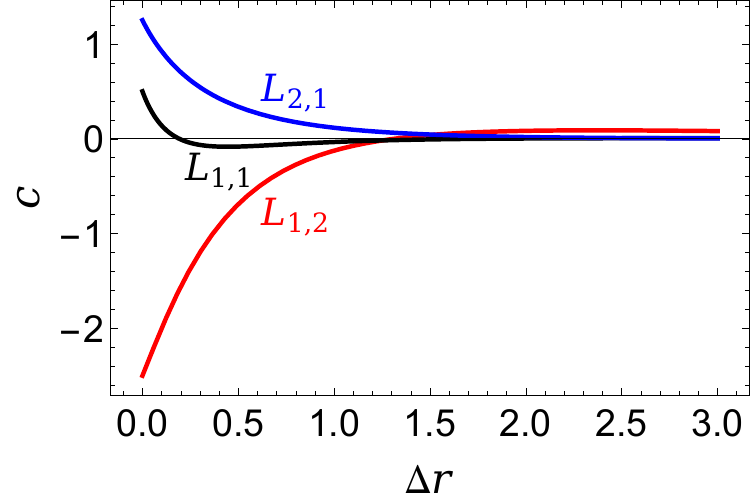}}
\end{center}
\caption{(Color online) (a) Behavior of (a) $\alpha_M$ and (b) axial torque $m$ and (c) parameter $c$ as function of excess radius $\Delta r$ for each of
$L_{2,1}$ (blue curve), $L_{1,1}$ (black curve) and $L_{1,2}$ (red curve)
sequences. In all three cases $\alpha_M$ initially increases with $\Delta r$
from a value $\alpha_M=0$, when $\Delta r=0$. The corresponding states are loops covering the equator described by perturbation theory. $\alpha_M$ reaches a maximum ($>\pi/2$) at a finite value of $\Delta r$. There is a critical
intermediate loop length at which $\alpha_M=\pi/2$, where the loop develops
vertical tangents. As the length is increased $\alpha_M$ falls asymptotically
to  $\pi/2$ corresponding to a loop stretched along the cylinder.  The axial
torque initially increases with $\Delta r$  ($m=(n/p)^2-2$ when $\Delta r=0$),
reaching a maximum, tending to zero as $\Delta r$ becomes large.  $c_0 = 3/2
-(n/p)^2$ when $\Delta r=0$ and $c\to 0$ when $\Delta r$ is large. Its vanish
correlates with $\alpha_m = \pi/2$.}
\label{Fig4}
\end{figure}

We now examine loops in their full non-linear glory. Their analysis is again
facilitated by exploiting the analogy with a particle in a periodic potential.
\vskip1pc \noindent
First, the constant $c$ can be cast in terms of the turning points of $U$ and
$m$: let $\alpha'=0$ at $\alpha=\pm \alpha_M$ in the quadrature Eq.
(\ref{CylQuad}) then
\begin{equation} \label{calphaMm}
c = -\frac{1}{2}\,\cos^4 \alpha_M - m \, \cos \alpha_M\,.
\end{equation}
For a circular loop this reproduces Eq. (\ref{m01}), relating $c$ and $m$
to lowest order in perturbations. $c$ vanishes if the curve is vertical on the
equator ($\alpha_M =\pi/2$) or when  $m = -1/2 \cos^3 \alpha_M$
($\alpha_M\neq\pi/2$). The relationship (\ref{calphaMm}) permits us to recast
the quadrature in terms of $\alpha_M$ and $m$: $\alpha'{}^2 = V(\alpha)^2$, where
\begin{equation} \label{quadnl}
 V(\alpha)^2\equiv 2 \left(c -U(\alpha)\right)=\left(\cos \alpha - \cos
\alpha_M\right) \, \left[ \left(\cos \alpha + \cos \alpha_M\right) \left(\cos^2
\alpha + \cos^2 \alpha_M\right) + 2 \, m\right]\,.
\end{equation}
Since $\pi \geq \alpha_M \geq \alpha \geq 0$, one has $\cos \alpha \geq \cos
\alpha_M \geq 0$, so $V^2 \geq0$ implies the lower bound on the axial torque,
$m \geq -2 \, \cos^3 \alpha_M$; thus there are no closed curves with $m < -2$.
Equation (\ref{calphaMm}) now implies the bounds on $c$, $c \leq 3/2 \, \cos^4
\alpha_M$ for $0< \alpha_M < \pi/2$ ($c > 3/2 \, \cos^4 \alpha_M$ for $\pi/2 <
\alpha_M < \pi$).  These bounds together imply the inequality $c \leq 3/2
(|m|/2)^{4/3}$.
\vskip1pc \noindent
For a given $p$-fold covering and $n$-fold dihedral symmetry, Eq.
(\ref{Deltaphi}) establishes the relationship between $m$ and $\alpha_M$.
Curiously,  this is independent of the length of the loop. Modulo this
condition, $\alpha_M$ or $m$ is determined as a function of $\Delta r$ using
Eq. (\ref{Deltar}). To trace  the trajectory of the loop  one integrates the
quadrature for a given $\Delta r$ to obtain $\alpha$ as a function of $\ell$;
the relations (\ref{sincosaalphacyl}) permit one to position the loop on the cylinder.
\vskip1pc \noindent
The functional relationship between $m$ and $\alpha_M$ is presented in Fig.
\ref{Fig3} for each of the three sequences, ${\rm L}_{2,1}$ (blue curve), ${\rm L}_{1,1}$ (black) and ${\rm L}_{1,2}$ (red).  While this may be the first
relationship to be established,  the information it conveys is not immediately
transparent.\footnote{Note, however, that it is consistent with the
perturbative behavior presented in Sect. \ref{sectpertsols}  for small
$\alpha_M$. Recall that for a $p$-fold covering of the equator, with
$\alpha=\alpha_M=0$, one has $m = (n/p)^2 -2$. (See Eqs. (\ref{mc0pert}).) The
leading correction is quadratic in $\alpha_M$, $m_2 =\frac{3}{4}\, \left(3 -
\frac{1}{2} \, (n/p)^2 \right) \,\alpha_{M}^2$.}
It is somewhat easier to understand the behavior of $m$ and $\alpha_M$ as
functions of $\Delta r$. In Figs. \ref{Fig4}(a), \ref{Fig4}(b) and \ref{Fig4}(c) $\alpha_M$, $m$ and $c$ are plotted as functions of loop length for each of the three sequences. We see that $m$ increases towards a maximum in each sequence. The  corresponding loops with maximum values of $m$ are illustrated in Figs. \ref{Fig5}(c), \ref{Fig6}(b) and \ref{Fig7}(d), respectively. Beyond this single maximum $m$ decreases monotonically to zero. The axial torque vanishes in all long loops. In each of the ${\rm L}_{2,1}$ and ${\rm L}_{1,1}$ sequences, $m$ changes sign from negative to positive at some finite $\Delta r$. In particular, there is a finite value of $\Delta r$ at which the axial torque vanishes. This does not, of course, mean that the total torque vanishes.\footnote{Likewise, $\mathbf{M}$ does not vanish in long loops. For while they are stretched along the cylinder, as we see, the asymptotic geometry is non trivial.} These states are tractable analytically and are discussed in detail in Appendix \ref{Sect:f0m0}. Surprisingly, one finds that $\alpha_M$  does not increase monotonically with loop length. It increases from its value $\alpha_M=0$ reaching a maximum value ($>\pi/2$) before returning asymptotically to  $\pi/2$ in long loops. At some critical length on the way up, $\alpha_M=\pi/2$, indicating that the loop develops vertical tangents on the equator. In these states the constant $c$ vanishes, which indicates that they are also special with respect to length. Loops with values of $\alpha_M >\pi/2$ develop overhangs on  each side of the equator.

\subsubsection{${\rm L}_{1,1}$ ground state}

\begin{figure}[htb]
\begin{center}
  \begin{tabular}{ccccccc}
  $\vcenter{\hbox{\includegraphics[scale=0.5]{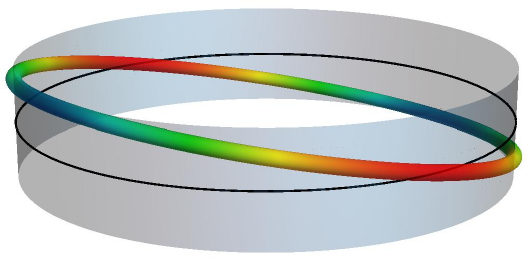}}}$ &
  $\vcenter{\hbox{\includegraphics[scale=0.4]{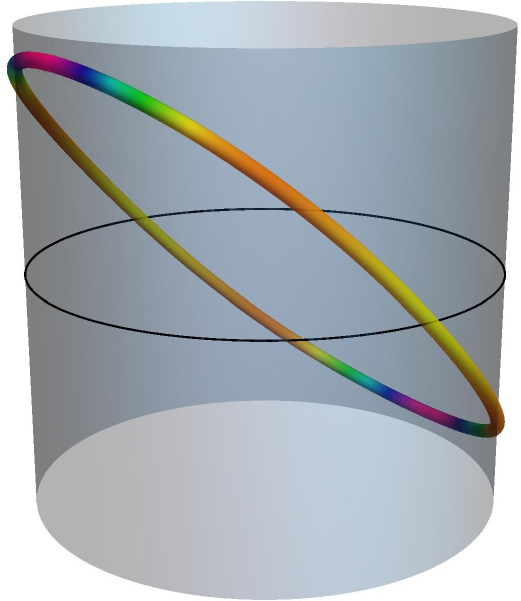}}}$ &
  \quad $\vcenter{\hbox{\includegraphics[scale=0.4]{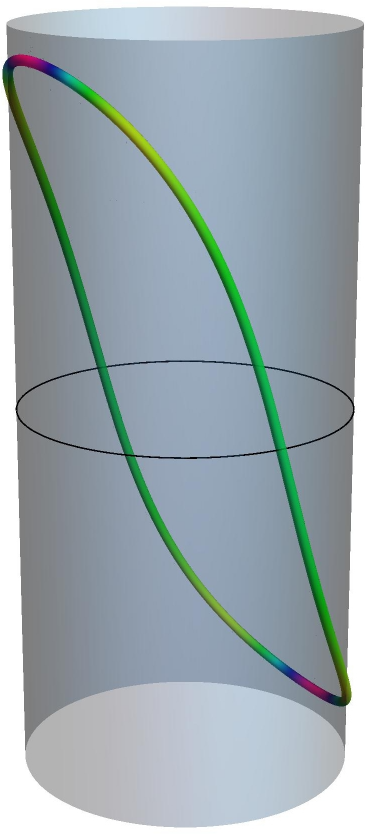}}}$ &
  \quad $\vcenter{\hbox{\includegraphics[scale=0.45]{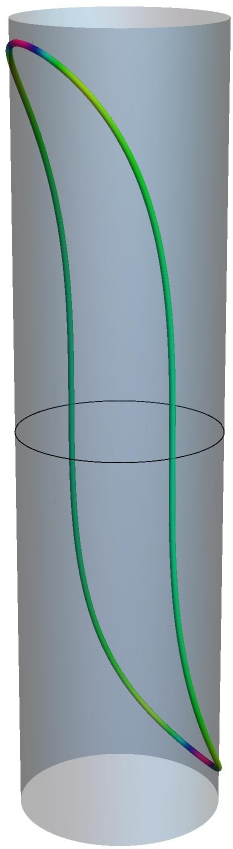}}}$ &
  \quad $\vcenter{\hbox{\includegraphics[scale=0.55]{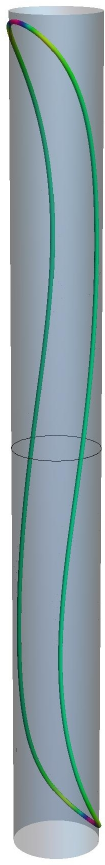}}}$ &
  \quad $\vcenter{\hbox{\includegraphics[scale=0.65]{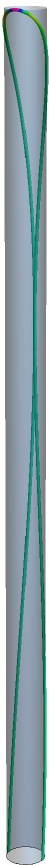}}}$ &
  \quad $\vcenter{\hbox{\includegraphics[scale=1]{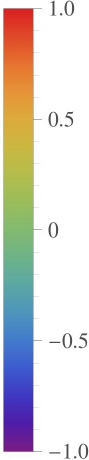}}}$\\
  {\small (a)} &
  {\small (b)} &
  \quad {\small (c)} &
  \quad {\small (d)} &
  \quad {\small (e)} &
  \quad {\small (f)} &
\end{tabular}
\end{center}
\caption{(Color online) ${\rm L}_{1,1}$ loop sequence. (a) $\Delta r=0.01$
($\alpha_M \approx \pi/16$), (b) $\Delta r=0.192$ ($\alpha_M \approx \pi/4$ and
$h_{B\, {\rm min}}$), (c) $\Delta r = 0.78$ ($\alpha_M \approx 2 \, \pi/5$ and
$m_{\rm max}$), (d) $\Delta r = 1.8$ ($\alpha_M = \pi/2$ and $h_{B\, {\rm
max}}$), (e) $\Delta r = 5.544$ ($\alpha_{M \, {\rm max}} \approx  0.548 \,
\pi$) and (f) $\Delta r = 27.819$ ($\alpha_M \approx 0.522 \, \pi$, first
self-contact occurs). In (f) only the upper half of the loop is displayed. The
normalized magnitude of the normal force $\lambda$ is color coded along the
loop in these figures.} \label{Fig5}
\end{figure}

Equilibrium loops completing one period in one trip around the equator are
represented in Fig. \ref{Fig5} focusing on the behavior as the length of the
loop is increased. If $\Delta r$ is small, the loop is a tilted ellipse
described accurately by perturbation theory about the equator. See Fig.
\ref{Fig5}(a). As the length increases, the eccentricity of this loop increases, obliging it to tilt away from the equatorial plane (Fig. \ref{Fig5}(b)); it also begins to bend out of the plane of the ellipse (Fig. \ref{Fig5}(c)) as inflections develop  at the equatorial crossings  (Fig. \ref{Fig5}(d)). This out of plane bending can be viewed  as a consequence of the mismatch between the curvature within the two hairpins  which follow  the parallel circles on the cylinder and the increasing linearity of the sections interpolating between them.  At some point the tangents on the equator become vertical with $\alpha_M=\pi/2$. With increased length,  overhangs appear and the loop develops lobes on either side of the equator. With increased $\Delta r$, the lobes continue to grow (Fig. \ref{Fig5}(e).\footnote{The angle $\alpha_M$, on the other hand, reaches a maximum, returning asymptotically to $\pi/2$.}  At some critical length the two lobes make self-contact at the back of the cylinder (Fig. \ref{Fig5}(f)). Meanwhile the two segments interpolating between the hairpins become increasingly vertical. The loop will eventually self-intersect. Very long loops consist of two hairpins connected by two very long self-intersecting almost vertical sections. Both the length of these sections and a number of self-intersections along them is approximately proportional to the length of the loop.

\subsubsection{${\rm L}_{1,2}$ excited states }

\begin{figure}[htb]
\begin{center}
  \begin{tabular}{ccc}
  $\vcenter{\hbox{\includegraphics[scale=0.5]{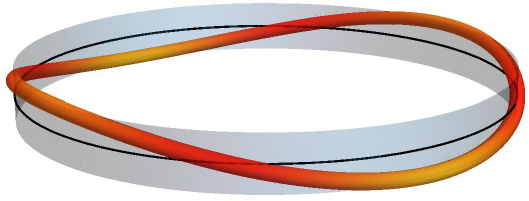}}}$ &
  $\vcenter{\hbox{\includegraphics[scale=0.4]{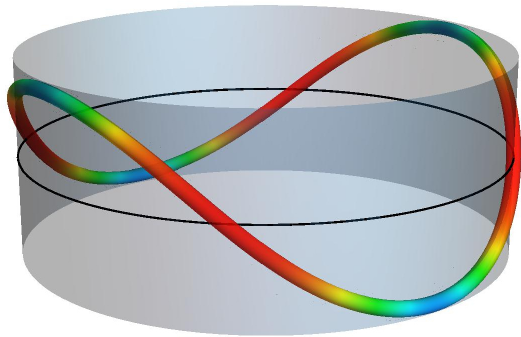}}}$ &
  $\vcenter{\hbox{\includegraphics[scale=0.4]{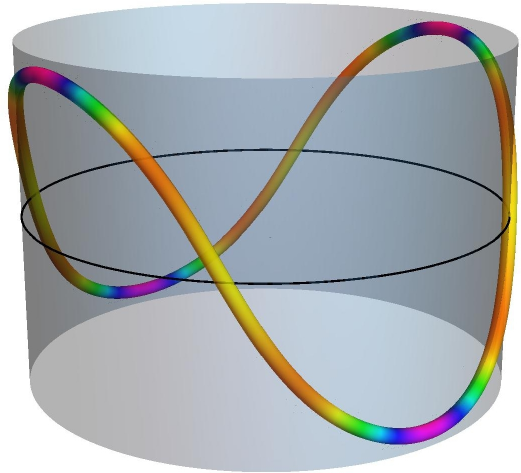}}}$ \\
  {\small (a)} &   {\small (b)} &  {\small (c)} \\
  $\vcenter{\hbox{\includegraphics[scale=0.45]{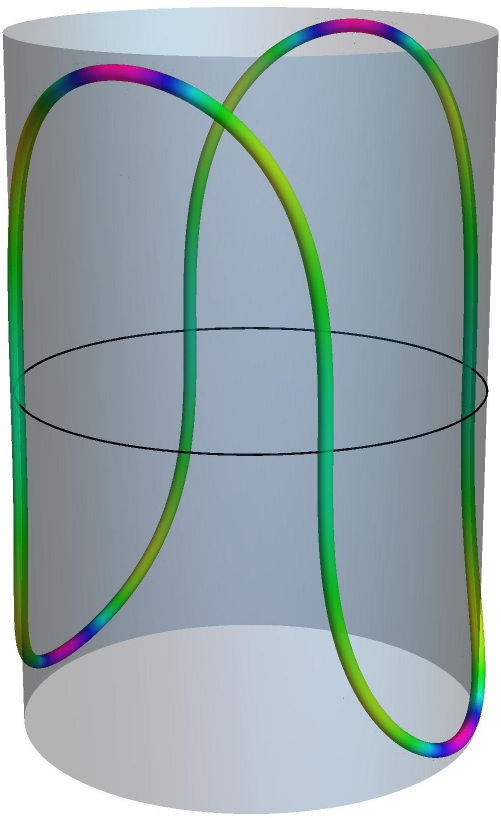}}}$ &
  $\vcenter{\hbox{\includegraphics[scale=0.55]{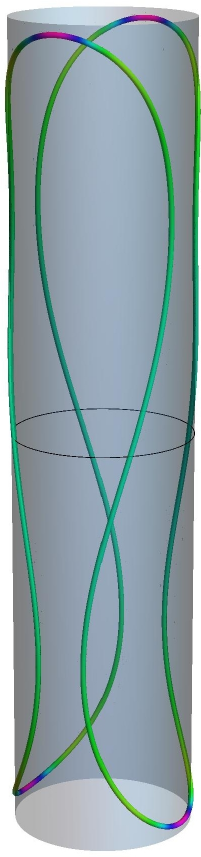}}}$ &
  $\vcenter{\hbox{\includegraphics[scale=0.65]{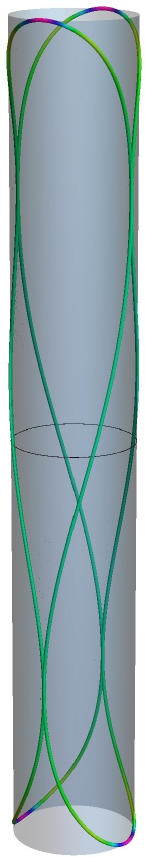}}}$ \\
    {\small (d)} &   {\small (e)} &  {\small (f)} \\
    &$\vcenter{\hbox{\includegraphics[scale=1]{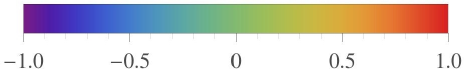}}}$&
\end{tabular}
\end{center}
\caption{(Color online) ${\rm L}_{1,2}$ loops sequence. (a) $\Delta r=0.01$
($\alpha_M \approx \pi/16$), (b) $\Delta r=0.13$ ($\alpha_M \approx 0.216 \pi$
and $m_{\rm max}$), (c) $\Delta r = 0.365$ ($\alpha_M = \pi/3$), (d) $\Delta r
= 1.313$ ($\alpha_M = \pi/2$ and $h_{B \, {\rm max}}$), (e) $\Delta r = 5.48$
($\alpha_{M \, {\rm max}} \approx  0.595 \, \pi$) and (f) $\Delta r = 8.59$ (
first self-contact occurs, $\alpha_M \approx 0.588 \, \pi$). The normalized
magnitude of the normal force $\lambda$ is color coded in these figures.}
\label{Fig6}
\end{figure}

Significant features of  loops undergoing two periods while they wrap the
cylinder once are represented in Fig. \ref{Fig6}. If the loop is short, it
consists of a small oscillation about the equator, (Fig. \ref{Fig6}(a)),  the
amplitude of the oscillation increasing with length  (Fig. \ref{Fig6}(b) and
(c)). When $\Delta r = 1.313$ the loop becomes vertical at the equator, (Fig.
\ref{Fig6}(d)). Lobes develop in longer loops, with $\alpha_M$ reaching its
maximum when $\Delta r = 5.48$ (Fig. \ref{Fig6}(e)). The first self-contact
occurs when $\Delta r = 8.59$ (Fig. \ref{Fig6}(f)). Self-intersections will
occur in longer loops. Asymptotically, states consist of four hairpin bends
interpolated by increasingly vertical mutually intersecting sections. One can
also think of this state as a folded $n=1$ state, with the folds themselves
forming hairpins. In general, states with $n>p$ will behave in a manner similar to those in this sequence.

\subsubsection{${\rm L}_{2,1}$ states}

\begin{figure}[htb]
\begin{center}
  \begin{tabular}{ccccccc}
  $\vcenter{\hbox{\includegraphics[scale=0.35]{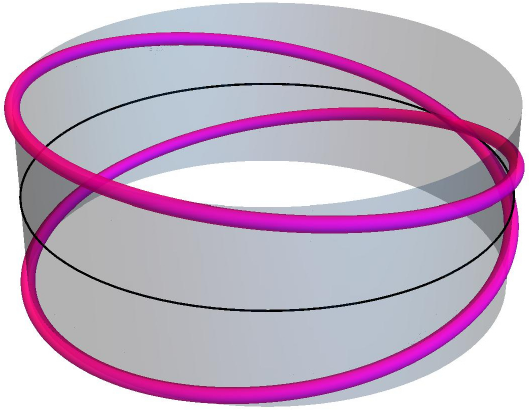}}}$ &
  \quad $\vcenter{\hbox{\includegraphics[scale=0.25]{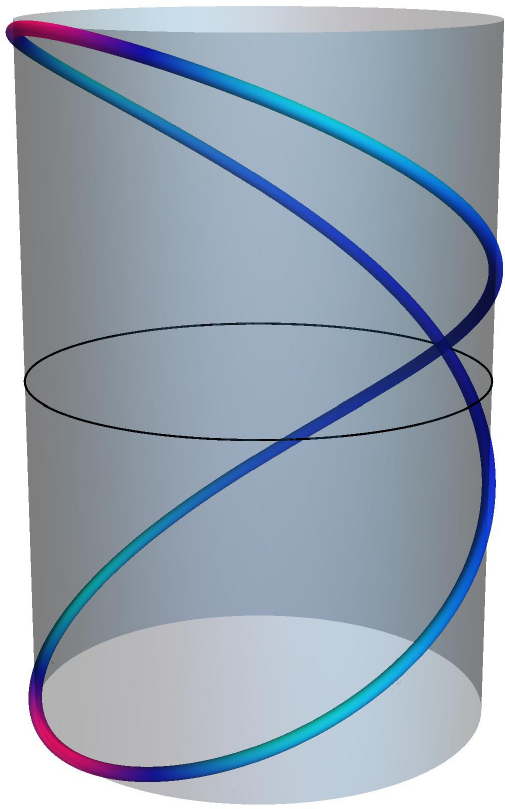}}}$ &
  \quad$\vcenter{\hbox{\includegraphics[scale=0.35]{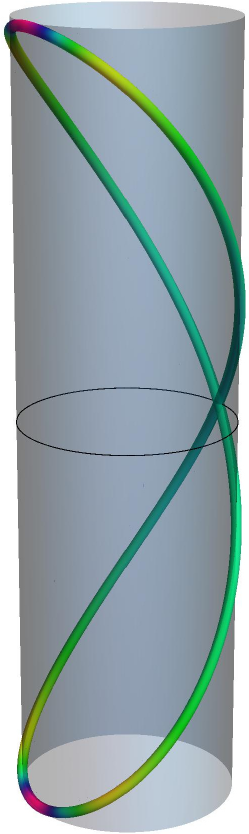}}}$ &
  \quad $\vcenter{\hbox{\includegraphics[scale=0.5]{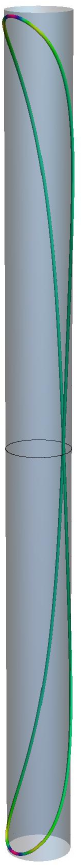}}}$ &
  \quad $\vcenter{\hbox{\includegraphics[scale=0.6]{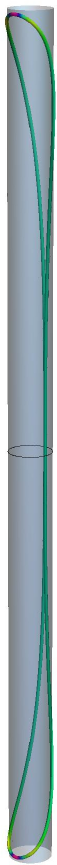}}}$ &
  \quad $\vcenter{\hbox{\includegraphics[scale=0.7]{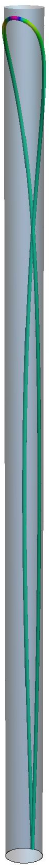}}}$ &
    \quad $\vcenter{\hbox{\includegraphics[scale=1]{Fig5g.pdf}}}$\\
  {\small (a)} &
  \quad{\small (b)} &
  \quad{\small (c)} &
  \quad {\small (d)} &
  \quad {\small (e)} &
  \quad {\small (f)} &
\end{tabular}
\end{center}
\caption{(Color online) Loops of sequence ${\rm L}_{2,1}$. (a) $\Delta r=0.02$
($\alpha_M \approx \pi/16$), (b) $\Delta r = 0.25$ ($\alpha_M \approx 0.194\,
\pi$),  (c) $\Delta r=1.149$ ($\alpha_M = \pi/3$),  (d) $\Delta r = 6.916$
($\alpha_M \approx 0.48 \pi$ and $m_{\rm max}$), (e) $\Delta r = 11.06$
($\alpha_M = \pi/2$ and $h_{B \, {\rm max}}$) and (f) $\Delta r = 27.548$
($\alpha_{M \, {\rm max}} \approx  0.51 \, \pi$). In (f) only the upper half is
displayed. The normalized magnitude of the normal force $\lambda$ is color
coded in these figures.} \label{Fig7}
\end{figure}

Non-trivial loops which wind about the cylinder more than once ($p>1$) are also
possible  whenever $r\ge p$. The  $L_{2,1}$  loop sequence is represented in
Fig. \ref{Fig7}. If the excess length is small the loop approximates  a double
covering of the equator,  adopting as the length is increased the shape of a
folded figure of eight wrapping the cylinder, with its self-intersection lying
on the equator. See Fig. \ref{Fig7}(a). (If $p=3$, the two intersections
migrate off the equator.) The crossing angle increases as the figure is
stretched. See Figs. \ref{Fig7}(b)-(d).  This angle becomes vertical when
$\Delta r = 11.06$, see Fig. \ref{Fig7}(e). $\alpha_M$ reaches a maximum value
$0.51 \, \pi$, at $\Delta r = 27.548$. See Fig. \ref{Fig7}(f). In longer loops,
additional self-intersections appears above and below the equator. In contrast
to the asymptotic behavior of the $L_{1,1}$ sequence, the two hairpins lie on
the same meridian.
\vskip1pc \noindent
The appearance of the self-intersections in these three sequences of loops can
be understood more easily by unfolding them as described in detail in Appendix
\ref{App:ComCylPEE}.

\subsubsection{Trajectories in the $c$-$m$ parameter space }

\begin{figure}
     \includegraphics[scale=0.6]{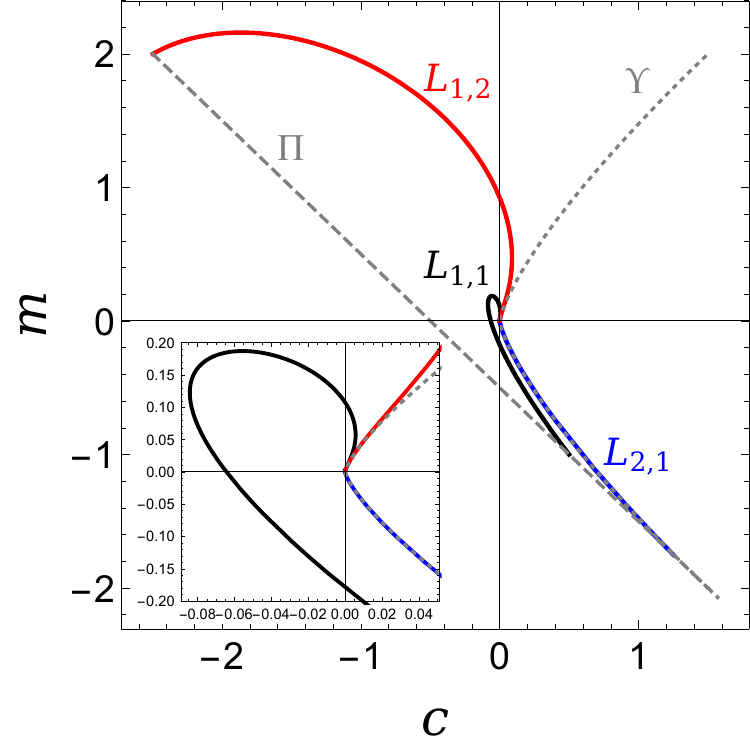}
    \caption{(Color online) Trajectories in the parameter space $c-m$ of
sequences ${\rm L}_{2,1}$ (blue curve), ${\rm L}_{1,1}$ (black curve) and ${\rm L}_{1,2}$ (red curve). Dashed line $\Pi$ represents the bound $m = -1/2 -c$, corresponding to a single or multiple covering of the equator, whereas dotted line $\Upsilon$ represents the bound $c = 3/2 (|m|/2)^{4/3}$ and correspond to loops covering the equator a very large number of times. The intersection of $\Pi$ and $\Upsilon$ occurs at the point $(3/2,-2)$. The inset shows the trajectories in the neighborhood of the origin, corresponding to loops with very large excess radius.} \label{Fig8}
\end{figure}

For completeness, we also represent the three sequences as trajectories in the
$c-m$ parameter space. These trajectories are bounded by two curves,

\begin{enumerate}
 \item The line $m = -1/2 -c$, denoted by $\Pi$, representing loops covering
the equator $p$ times, is indicated by the gray dashed line in Fig. \ref{Fig8}.
 \item The curve $c = 3/2 (|m|/2)^{4/3}$,  denoted $\Upsilon$ representing
loops saturating the inequality between $c$ and $m$, is indicated by the  gray
dotted curve in Fig. \ref{Fig8}.
\end{enumerate}
The point $(3/2,-2)$ at the intersection of $\Pi$ and $\Upsilon$, is a limit
point, corresponding to loops  winding the equator an ever increasing number of
times ($p \rightarrow \infty$),  with a finite number of periods, so that $c_0
\rightarrow 3/2$ and $m_0 \rightarrow -2$.
\vskip1pc \noindent
All three trajectories originate on $\Pi$ and terminate at the origin.
Infinitely long loops have vanishing $m$ and $c$.
In this limit the fixed length constraint gets relaxed as $c$ becomes
vanishingly small.
\vskip1pc \noindent
The trajectory for ${\rm L}_{1,1}$ (${\rm L}_{2,1}$) indicated by the black
(blue) curve in Fig. \ref{Fig8} begins at the point $(1/2, -1)$ ($(5/4, -7/4)$)
spiraling towards the origin.  Both trajectories cross the line $c=0$ three
times; twice with a finite value of $m$, one with an acute angle and another
with vertical tangents at the equator,  and again at the origin.
\vskip1pc \noindent
Likewise, the trajectory for ${\rm L}_{1,2}$ begins on $\Pi$ at $(-5/2, 2)$ but it forms an arc,  crossing the line $c=0$ twice, once where the loop has
vertical tangents and again at the origin.
\vskip1pc \noindent
In general, any sequence with $n>p$ begins on $\Pi$ to the left of ${\rm
L}_{1,1}$, whereas sequences with $n<p$ begin on the right, all sequences tend
asymptotically to the origin. Sequences with $c>0$ ($c<0$) or $n/p <
\sqrt{3/2}$ ($n/p > \sqrt{3/2}$) will cross the line $c = 0$ three times
(twice) and thus will have spiral-like (arc-like) trajectories as in sequence
${\rm L}_{2,1}$ (${\rm L}_{1,2}$).

\subsubsection{Total energy}

\begin{figure}
 \begin{center}
    \includegraphics[scale=0.675]{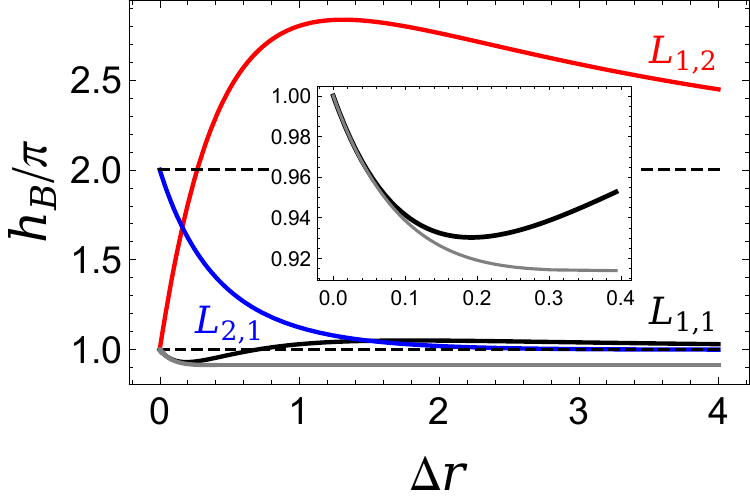}
 \end{center}
    \caption{(Color online) Total energy of loops belonging to the three
sequences ${\rm L}_{2,1}$ (blue), ${\rm L}_{1,1}$ (black) and ${\rm L}_{1,2}$
(red), as well as the energy of the vertical loop touching the cylinder along
its sides (gray). The initial value of the total energy of the ${\rm L}_{p,n}$
sequence is $h_B = \pi p$, corresponding to a $p$-fold covering of the equator, whereas the energy of loops with very large excess radius tend asymptotically
to the value $h_B = \pi n $, indicated  with dashed lines.} \label{Fig9}
\end{figure}

Loops in the ${\rm L}_{p,n}$ sequence  originate in $p$-fold coverings of the
equator. Thus their initial energy is $h_B = \pi \, p$, independent of $n$. See Fig. \ref{Fig9}. If the excess radius $\Delta r$ is small, loop states in the
${\rm L}_{1,1}$ sequence are approximately elliptic; their energy is thus close to that of their planar vertical  counterparts presented in Appendix
\ref{SectConfPEE}, as revealed by the initial coincidence of the solid black
and gray curves, respectively, in Fig. \ref{Fig9}. However,  in this sequence the energy reaches a global minimum when $\Delta r = 0.192$, corresponding to the state illustrated in Fig. \ref{Fig5}(b), increasing towards a global maximum value $h_B = 1.042 \, \pi $ when $\Delta r = 1.80$ (the state with $\alpha_M = \pi/2$ and $c=0$), after which it decreases asymptotically towards $h_B = \pi$. Long loops and short ones have the same energy! By contrast, the energy of planar vertical loops decreases monotonically to a minimum value of $0.914 \pi$ when $\Delta r = 0.393$ which is  when the curvature in the  free loop vanishes where it makes contact with the wall. Thereafter it remains constant because any additional length gets directed into the straight line segments. See the inset in Fig. \ref{Fig9}. We see that vertical loops always have lower energy than any bound counterpart: the former thus provides the ground state of interior bound loops allowed to unbind.
\vskip1pc \noindent
The energy of excited states in the ${\rm L}_{1,2}$ sequence is indicated by
the red line in Fig. \ref{Fig9}. While  it is initially degenerate with the
${\rm L}_{1,1}$ sequence,  with energy $h_B = \pi$, unlike the latter it
increases with the excess radius from the beginning, reaching a maximum at
$\Delta r = 1.313$ (also when $\alpha_M = \pi/2$), thereafter decreasing
asymptotically to the value $h_B = 2\, \pi$. The energy of the double winded
${\rm L}_{2,1}$ sequence  is indicated by the blue curve in Fig. \ref{Fig9}.
It starts at $h_B = 2\,\pi$; unlike the energy of the ${\rm L}_{1,1}$ sequence
which only decreases initially, the energy decreases monotonically to $h_B =
\pi$, the same asymptotic energy as a loop in the $L_{1,1}$ sequence.
\vskip1pc \noindent
In general, for  ${\rm L}_{p,n}$ sequences with $n>p$, the maximum of $h_B$
occurs in states with vertical tangents on the equator and finite vertical
torque, i.e., $\alpha_M = \pi/2$ and $m \neq 0$. The corresponding states in the ${\rm L}_{1,1}$ and ${\rm L}_{1,2}$ sequences are displayed in Figs.
\ref{Fig5}(d) and \ref{Fig7}(e). For $n<p$ the initial states possess the
maximum energy within the sequence. Since loops with very large excess radius
are essentially straight almost everywhere, the principal contributions to the
total energy will be from the $2\,n$ curved hairpins, each with approximately
the same energy as a semicircle, $h_B = \pi/2$.  This explains the coincidental simplicity of the asymptotic energy in the three sequences: for all ${\rm
L}_{p,n}$, the energy $h_B \rightarrow \pi \, n$  (from  above) as $\Delta r
\rightarrow \infty$. It is independent of the topology of the loop.

\subsubsection{Transmitted normal force}

\begin{figure}
 \begin{center}
 \subfigure[]{\includegraphics[scale=0.65]{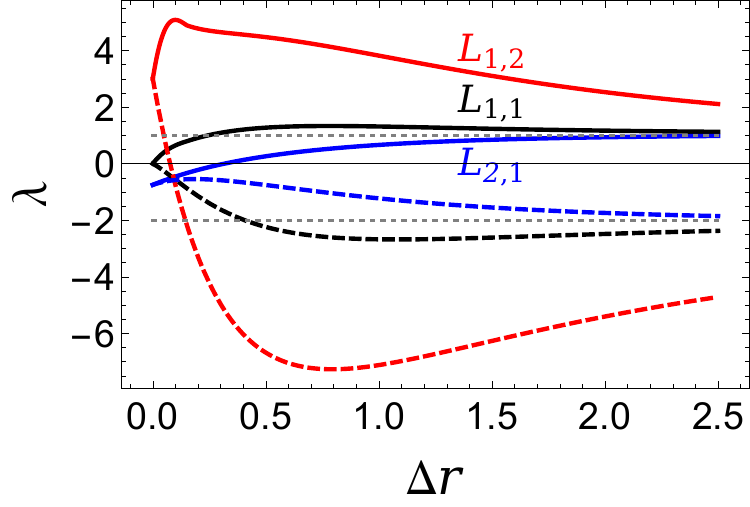}} \hfil
\subfigure[]{\includegraphics[scale=0.65]{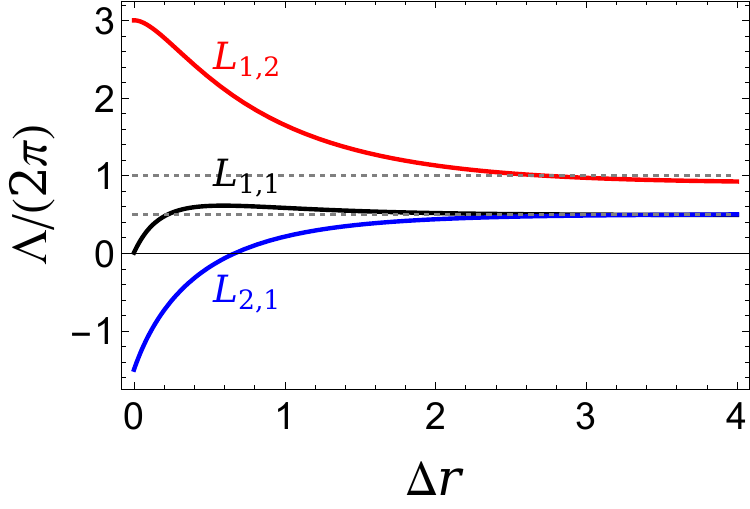}}
 \end{center}
 \caption{(Color online) (a) Maximum (solid curves) and minimum (dashed curves)
of the magnitude of the normal force $\lambda$ for sequences ${\rm L}_{2,1}$ (blue
curve), ${\rm L}_{1,1}$ (black curve) and ${\rm L}_{1,2}$ (red curve). (b) Total normal
force of the same sequences.} \label{Fig10}
\end{figure}

The magnitude of the normal force $\bar{\lambda}$, scaled to lie between $-1$
and $1$, is represented in color along the loops in Figs. \ref{Fig5},
\ref{Fig6} and \ref{Fig7}, indicating  clearly the regions where they are
pushing or pulling the cylinder: red (purple) represents regions where the push (pull) is maximum,\footnote{Recall that this convention is  adapted to the
viewpoint of an interior loop; if the loops is outside, push and pull are
reversed.} whereas green represent  regions in which the transmitted force is
low. In general, it is observed that the maximum push occurs at the equatorial
crossings, whereas the minimum (maximum pull if negative) occurs at the
hairpins. The maximum (solid curves) and minimum (dashed curves) values of
$\bar{\lambda}$ are plotted as functions of the excess radius $\Delta r$ in Fig. \ref{Fig10}(a) for each of the three ${\rm L}_{1,1}$ (black curve), ${\rm
L}_{1,2}$ (red curve) and ${\rm L}_{2,1}$ (blue curve) sequences.
\vskip1pc \noindent
The perturbative result, Eq. (\ref{lambdapert}), indicates that the initial state of a sequence ${\rm L}_{p,n}$ exerts a normal force of magnitude
\begin{equation}
\bar{\lambda}_{0} = (n/p)^2-1\,.
\end{equation}
Regardless of $n$ and $p$, the maximum and minimum values of $\bar{\lambda}$
tend to $\bar{\lambda}_{max} \rightarrow 1$ approximately and
$\bar{\lambda}_{min} \rightarrow -2$ exactly as $\Delta r \rightarrow \infty$.
In fact, in this limit Eq. (\ref{eq:Fcyl1mf}) reads $\bar{\lambda} \approx 2 \, \cos^4 \alpha \left(5-6 \cos^2 \alpha \right)$, with minimum at $\alpha=0$ and
a maximum when $\cos^2 \alpha= 5/9$, so that $\bar{\lambda}_{max} = 250/243$,
which is $1$ for all practical purposes.
\vskip1pc \noindent
For completeness, we present the total transmitted force $\Lambda = \oint {\rm
d}s \, \bar{\lambda}$, which indicates, whether, on average the loop is pushing or pulling on the cylinder.  It is plotted in Fig. \ref{Fig10}(b) in each of
the three sequences. Since the $p$-fold coverings of the equator exert a
constant normal force, the initial total force is simply $\Lambda_0 = 2 \pi \,
p \, \bar{\lambda}_{0}$. It is found that $\Lambda$ tends asymptotically
to $\Lambda_\infty \rightarrow \pi \, n$ from below as $\Delta r \rightarrow
\infty$. Whereas enclosed bound loops both push or pull on the cylinder,
depending on the sign of $\bar{\lambda}_{0}$, ultimately all such loops
end up pushing the cylinder.
\vskip1pc \noindent
In the  ${\rm L}_{1,1}$ sequence, the single covering of the equator does not
initially exert any force, $\bar{\lambda}_0 = 0$.  As $\Delta r$ increases,
however, second order corrections kick in with $\bar{\lambda}$ alternating from positive to negative along the loop. This behavior persists with increasing
$\Delta r$; see Fig. \ref{Fig5} and the black curves in Fig. \ref{Fig10}(a).
The negative pulling force on an inner bound loop is increasingly localized in
the neighborhood of the tips. We saw earlier that the energy of this state is
always higher than that of the  planar vertical loop of equal length spanning
the interior. Importantly, however, the averaged force is always positive.
While the energy would suggest that the ${\rm L}_{1,1}$ ground state is always
unstable,  the sign of the contact force suggests otherwise. What appears to
occur is that the loop unbinds from the cylinder in the neighborhoods of the
tips. If it is short it will rotate into the vertical. As the length increases,  however, the lobes pushing on the cylinder snag the loop providing a potential barrier obstructing the passage towards  the planar vertical loop. A detailed
perturbative study, beyond the scope of this paper, is required to settle the
issue.
\vskip1pc \noindent
Along loops in the ${\rm L}_{1,n}$ sequences, initially $\bar{\lambda}$ is
uniformly positive (see Fig. \ref{Fig6}(a)), indicating that an Euler type
instability is associated with their formation as interior bound states.
However this push on the cylinder persists only for short loops; once $\Delta r > 7.65 {\rm x} 10^{-2}$ for $n=1$ (see the red dashed line in Fig. \ref{Fig10}(a)), the minimum of $\bar{\lambda}$ turns negative within the
regions near the tips, indicated in Fig. \ref{Fig6}(b). In the case of an
interior bound state,  one would expect the loop to unbind in the neighborhood
of its tips so as avoids these regions.
\vskip1pc \noindent
An exterior bound loop with $p=1$, free to unbind will completely unbind,
reforming as a circular loop.
As we have already glimpsed, loops in the  ${\rm L}_{2,1}$ sequence behave very differently: now $\bar{\lambda}$ is initially uniformly negative.  While an
inner bound loop in this state is manifestly unstable initially, its
exterior bound counterpart is stable. However when $\Delta r > 0.317$, the
maximum of $\bar{\lambda}$ becomes positive (c.f. the blue solid line in Fig.
\ref{Fig10}(a)) in the regions surrounding the center of the hairpins; see Fig. \ref{Fig7}. Beyond some higher value of $\Delta r$ the average force becomes
positive.  This suggests that a sufficiently long interior loop may be stable
modulo shortcuts.
\vskip1pc \noindent
As discussed in the perturbative context, doubly-wound exterior loops are more
interesting because such loops will always constrict the cylinder. If the loop is short, it will constrict the cylinder everywhere, albeit not in an axially
symmetric way. If longer, the contact force will favor the unbinding of the
exterior bound loop in the neighborhood of the crossings. However, such loops
will remain  attached to the cylinder at its hairpins applying a force on the
cylinder at the points of contact. In any case the cylinder provides a
topological obstruction preventing the doubly wound elastic loop from
unraveling into a singly-wound circular loop.  This happy conspiracy permitted
by the topology has no analogue on a sphere.

\section{Discussion and conclusions} \label{Sect:Conclusions}

We have examined the equilibrium shapes adopted by an elastic loop either
confined by a cylinder or constricting it. Our initial motivation was to
contrast the behavior of loops confined within a cylinder with that of their spherical counterparts, and specifically to explore how these loops respond to the changed topology. Loops may wind a cylinder any integer number of times. They may oscillate one or more times as they do this. Whereas the winding has no topological significance for a loop that is free to unbind within the cylinder, it very much does if it is bound to the outer surface.
\vskip1pc\noindent
On a sphere, the ground state of a confined loop with a radius exceeding the
spherical radius is attached to the sphere. No additional adhesive forces are
necessary. The loop pushes against the sphere and it will remain bound
everywhere even if free to detach. Instabilities of excited states occur
through rearrangements on the sphere itself. When there are topological
obstructions, they may be bypassed by stepping into the interior. Elastic
loops, however, simply do not bind to the exterior surface of a sphere. What
occurs on a cylinder is a lot more interesting.
\vskip1pc\noindent
On a cylinder, one needs to distinguish between stability on the cylinder and
off it. We have considered, in detail, the behavior of states bound to the
cylinder. If the loop is short there is a single bound ground state for each
winding $p$, the state with $n=1$. All states with $n\ge 2$ will be unstable if the loop is short. However, as loops grow in length the sections connecting
the troughs and crests straighten out along the axis; if the loop oscillates
more than once, the oscillations morph into folded hairpins. All loops grow by
extending indefinitely in the axial direction, which permits them to relax
their axial torque.  One can show that for each $n \ge 2$ there will be a
critical length above which the bound folded loop is stabilized.  There will be a crossover from instability to stability associated with the exploration of
the long (or non-compact)  direction. This behavior has no analog on spheres.
\vskip1pc\noindent
If states are free to unbind,  the behavior inside and out are very different;
both are interesting. Unlike loops binding to spheres,  the contact forces  are not positive everywhere.  Indeed, they are always negative in the neighborhood
of the crests and troughs of the oscillations.  Interior bound states would be
expected to unbind within these regions, lowering the loop curvature and, as a
consequence, lowering the loop energy. The ground state itself with $p=1$,
$n=1$ will also unbind, and if sufficiently short, will be unstable with
respect to deformations rotating  it off the surface into the confined ground
state: a planar loop compressed along its sides. Experimenting with a loop in a cylindrical wastepaper basket or pipe confirms this prediction. Longer
partially bound states may become stable beyond some critical length.  Folded
hairpins will certainly be stable modulo minor adjustments. A stability
analysis accommodating normal perturbations is necessary to settle these
issues, but it is not going to be simple.
\vskip1pc\noindent
We saw that states with $p>1$ and $n=1$ would tend to unbind  if below some
critical length. This is because the contact forces are inward everywhere. As
these states get longer the force turns to push except at the tips so that the
state would be expected to stabilize modulo the same sort of corner-cutting we
have described. These states are, however, far more interesting  placed outside the cylinder.
\vskip1pc\noindent
An exterior loop wrapping a cylinder once will always unbind in the absence of
an adhesive force holding it there, reconfiguring itself as a simple circular
loop. However, one that wraps the cylinder more than once is constrained by the topology to remain in contact with the cylinder, completely initially and
partially--in the neighborhood of its tips--beyond some critical length. In
either case it will always apply a compressive force on the cylinder in these
neighborhoods. This force generally is not axially symmetric. This behavior can
also be confirmed in a broom cupboard experiment by coiling an elastic loop
into a figure-eight and wrapping a broom handle: whereas a short loop will
hug the handle completely, a long one will detach everywhere except at four
points. This experiment also reveals that the detached state does not possess
the symmetry of its bound counterpart. Indeed, if the loop is coiled three or
more times, it is clear that the sequence of crossing (over or below) is not
unique and the detached state will depend on the particular sequence. This
mechanism may be relevant to an understanding of the constriction of membrane
necks in dynamin mediated membrane vesiculation. It should be remarked that we
never anticipated the possibility that loops wrapping a cylinder could bind
without any additional agency, never mind, finding that they provide a possible mechanism for neck constriction.
\vskip1pc\noindent
A membrane neck is, of course, better modeled as a catenoid. Modeling the
constriction of a catenoid is technically more challenging because, along with
the surrender of the translational symmetry along the cylinder along its axis,
one also cedes  the  quadrature.
\vskip1pc \noindent
An obvious direction for future work is to study open elastic spirals on
cylinders, catenoids and other axially symmetric geometries with free
boundaries. Spiral geometries represent the behavior of chains of assembled
proteins such as dynamin. However, unless physically unrealistic BCs are introduced, the simple elastic energy considered here will not constrict the geometry. A constricting spiral will require an environmental bias, such as a spontaneous normal curvature, if it is to apply a force. If it is to spiral, it will require a constraint or bias on the twist which translates into a constraint or bias on the geodesic torsion.  This still leaves us with a long way to go  before we are in a position to treat a spiral binding to a fluid membrane taking their mutual interaction into account. The consolation is that there is interesting physics to be picked up along the way.

\section*{Acknowledgements}

We have benefited from conversations with Bojan Bo\v zic, Martin M. M\"uller,
M\'onica Olvera de la Cruz and Sa\v sa Svetina. We are grateful to Yair
Gutierrez for discussions at an early stage of this work. P.V.M. acknowledges
support from CONACYT Postdoctoral Grant 205393, as well as financial support by CMU. J.G. is partially supported by CONACYT under Grant 180901. M.D. and Z.M. are supported in part by the National Science Foundation under Grant CHE-1464926.

\begin{appendix}

\section{Axisymmetric Surfaces} \label{Appaxisymsurf}

To describe an axisymmetric surface, it is convenient to introduce cylindrical
coordinates $(R,\varphi, Z)$, adapted to the symmetry with tangent vectors
\begin{equation}
\hat{\bm \rho} =  \left( \cos \varphi,\, \sin \varphi,\, 0 \right)\,, \qquad
\hat{\bm \varphi} = \left( - \sin\varphi, \cos \varphi, \,0 \right) \,, \qquad
\hat{\bf z} = (0,0,1) \,. \\
\end{equation}
The surface is now parametrized by arc-length $\ell$ along a meridian and the
azimuthal angle $\varphi$ along the parallel,
\begin{equation} \label{Xlphiaxisym}
\mathbf{X}(\ell,\varphi) = R(\ell)\hat{\bm \rho} + Z(\ell) \hat{\bf z}\,,
\end{equation}
where $\dot{R}(\ell)^2 + \dot{Z}(\ell)^2 = 1$ and the dot represents derivation with respect to $\ell$.
The two tangent vectors adapted to this parametrization are
\begin{equation}
\mathbf{e}_\ell =  \dot{R} \hat{\bm \rho} + \dot{Z} \hat{\bf z}\,, \qquad
\mathbf{e}_\varphi = R \hat{\bm \varphi}\,.
\end{equation}
Thus, the line element on the surface is given by
${\rm d} s^2 = {\rm d} \ell^2 + R(\ell)^2 \, {\rm d} \varphi^2$,
and the metric tensor  assumes the form
\begin{equation}
g_{ab} = \left(
\begin{array}{cc}
1&0\\
0&R^2
\end{array}
\right)\,.
\end{equation}
The outward unit normal vector,  $\mathbf{n} =g^{-1/2}\, \mathbf{e}_\varphi \times \mathbf{e}_\ell$, is
\begin{equation} \label{normalaxisym}
\mathbf{n} = \dot{Z} \hat{\bm \rho} - \dot{R} \hat{\bf z}\,;
\end{equation}
the extrinsic curvature tensor is diagonal,
\begin{equation} \label{KabSAxs}
K_{ab} = \left(
\begin{array}{cc}
 - \ddot{R}/\dot{Z} & 0 \\
 0 & R \dot{Z}
\end{array}
\right)\,.
\end{equation}
The eigenvectors of the shape operator $K^a{}_b= g^{ac} K_{cb}$ (its principal
directions) lie along meridians and parallels, with corresponding eigenvalues
given by
 \begin{equation} \label{princurvaxisymm}
\kappa_\perp = -\ddot{R}/\dot{Z} \,, \qquad \kappa_\parallel = \dot{Z}/R\,.
\end{equation}
These are the curvatures of these curves. The two symmetric invariants of the
shape operator are  the mean and Gaussian curvatures $K = \kappa_\perp +
\kappa_\parallel$ and $K_G = \kappa_\perp \, \kappa_\parallel$.

\section{Hamiltonian approach to cylindrical confinement} \label{Appcylhamform}

Here we provide a direct derivation of the ``second'' integral of the
EL equation describing the confinement of a semi-flexible polymer
within a cylinder. The ``Lagrangian'' density $\mathscr{L}$ is constructed in
terms of the three coordinates $(\varphi,z,\alpha)$\footnote{As before, here we scale lengths with the cylinder radius.} characterizing the curve $\Gamma$.
These variables are not independent; they are related by Eqs.
(\ref{sincosaalpha}). It will be necessary to implement these constraints using two local Lagrange multipliers $\lambda_\varphi$ and $\lambda_z$. One thus
constructs the following effective energy
\begin{equation}
 \mathscr{L}[\alpha, \varphi,z,\alpha',\varphi',z',\lambda_\varphi,\lambda_z] =
\int {\rm d}s \left(\frac{(\alpha')^2}{2}+\frac{\cos^4 \alpha}{2}
+\lambda_\varphi (\varphi'-\cos \alpha) + \lambda_z (z'-\sin \alpha)\right)\,.
\end{equation}
The momentum densities conjugate to the coordinates $(\varphi,z,\alpha)$, are
given by
\begin{equation} \label{cylmomenta}
P_\varphi = \lambda_\varphi\,, \qquad P_z = \lambda_z\,, \qquad P_\alpha =
\alpha'\,.
\end{equation}
Since $\varphi$ and $z$ are cyclic coordinates (because of the rotational and
translational symmetries) $P_\varphi$ and $P_z$ are conserved. Also, $L$ does
not depend explicitly on arc-length. Thus, its Legendre transformation ${\cal H}= \varphi' P_\varphi + z' P_z + \alpha' P_\alpha  -{\cal L}$ is constant. Evaluating it, one obtains
\begin{equation}
{\cal H}= \frac {1}{2} P_\alpha^2-\frac{\cos^4 \alpha}{2} + P_z \sin \alpha +
P_\varphi \cos \alpha\,.
\end{equation}
This reproduces the ``second'' integral Eq. (\ref{2ndintcylconf}) with the
identifications, ${\cal H}=c$, $f=-P_z$ and $m = - P_\varphi$.
\vskip1pc \noindent
Hamilton's equations for $\alpha$ are
 \begin{equation}
 \alpha' = \frac{\partial {\cal  H}}{\partial P_\alpha} = P_\alpha\,, \quad
P'_\alpha = -\frac{\partial {\cal H}}{\partial \alpha} = -2 \, \cos^3 \alpha \,
\sin \alpha + P_\varphi \, \sin \alpha - P_z \, \cos \alpha \,.
 \end{equation}
To reconstruct the curve one solves these equations with initial conditions
$\alpha(0)=\alpha_0$, $P_\alpha(0)=P_{\alpha\,0}$ and fixed $P_\varphi$ and
$P_z$. The azimuthal and height functions are obtained from integration of the
two Hamilton equations
\begin{equation}
  \varphi' = \frac{\partial {\cal H}}{\partial P_\varphi} = \cos \alpha \,,
\quad  z' = \frac{\partial {\cal H}}{\partial P_z} = \sin  \alpha\,.
\end{equation}
Without loss of generality one can use the initial conditions $\varphi(0)=0$
and  $z(0)=0$. The three constants $P_z$, $P_\varphi$ and ${\cal H}$ are
determined from boundary or periodicity conditions.

\section{Confined planar curve} \label{SectConfPEE}

\begin{figure}[htb]
\begin{center}
\begin{tabular}{ccc}
$\vcenter{\hbox{\includegraphics[scale=0.62]{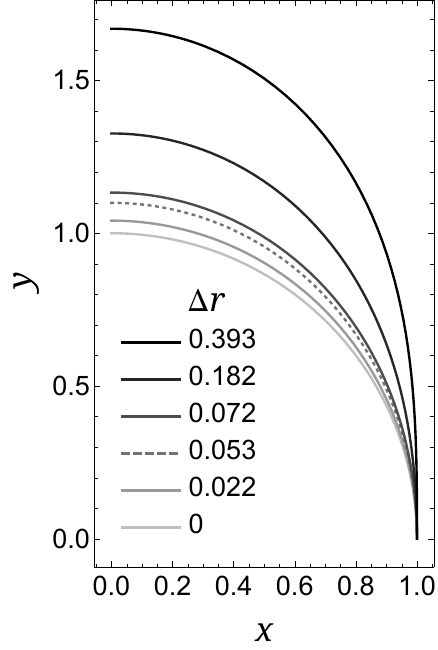}}}$ &
$\vcenter{\hbox{\includegraphics[scale=0.62]{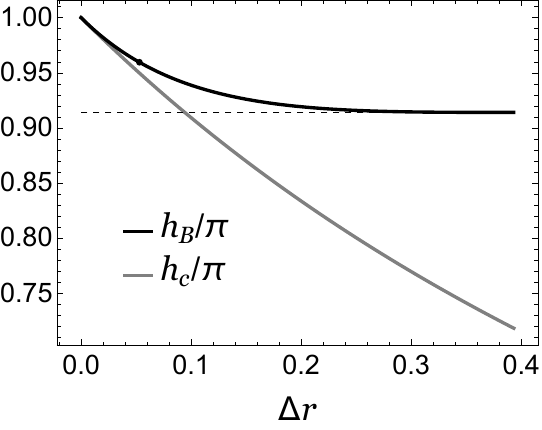}}}$ &
$\vcenter{\hbox{\includegraphics[scale=0.44]{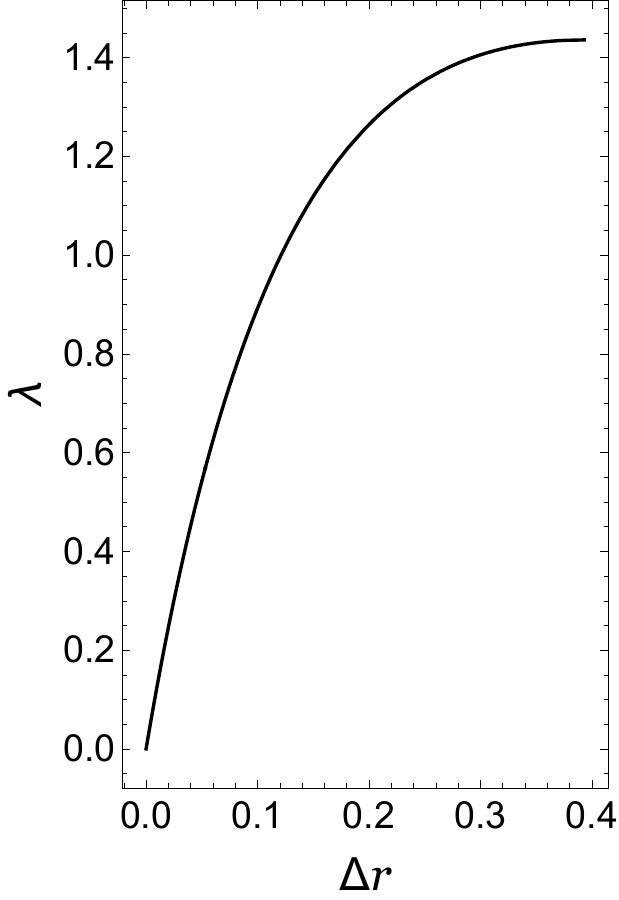}}}$\\
  {\small (a)} & {\small (b)} & {\small (c)}
\end{tabular}
\end{center}
\caption{ (a) One quarter of vertical elastic loops of different radii $R$
confined inside a cylinder of $R_0$ ($\Delta r=R/R_0-1$). The loops have
vertical tangent at two points where they touch the cylinder. The shortest
curve is the circle with radius equal to that of the cylinder ($\Delta r=0$). In the interval $0 < \Delta r < 0.053$ the loop is composed of arcs of orbitlike elasticas (light gray curve), when $ \Delta r = 0.053$ of a segment of the borderline elastica (dashed curve) and after which is formed by wavelike elastica segments (dark gray curves), reaching a limit shape at $\Delta r=
0.393$ (black curve), such that beyond this point the excess length will be
located at the contact regions, so the vertical intermediate region will
increase. (b) Bending energy of the confined loops as a function of their
reduced radius. The black dot represent the energy of the loop formed by
borderline elastica segments; curves with smaller (greater) radius consist of
orbitlike (wavelike) elasticas. $H_B$ starts at a value of $\pi/R_0$ and decreases monotonically (black curve)--but much slower than the bending energy of a free loop (gray line)-- as the loop radius increases, reaching a limit value of $0.914 \, \pi$ for the curve with $\Delta r = 0.393$, after which it remains constant, regardless of the loop length. (c) Magnitude of the normal force transmitted to the cylinder at the points where contact is made.} \label{Fig13}
\end{figure}

Let us consider the confinement of a closed curve of radius $R$, lying
vertically inside a cylinder of radius $R_0$, touching the cylinder
tangentially on its two sides. We have claimed that this state has lower energy than its bound counterpart.  In the absence of additional adhesive forces,  the bound interior ground state--if sufficiently short--would be expected to be
unstable with respect to deformation into the cylinder rotating it into this
state. Since the curve is planar, it is described by the EL equation \cite{Langer1984, SingerSantiago2008}
\begin{equation}  \label{planarELeq}
\varepsilon_{\bf N} = \kappa'' + \kappa \left(\frac{\kappa^2}{2} - c\right) = 0\,.
\end{equation}
The translational invariance of the bending energy permits one to identify a
quadrature
 \begin{equation} \label{planarquadrature}
 (\kappa')^2+\left(\frac{\kappa^2}{2} - c\right)^2 = F^2\,,
\end{equation}
where $F$ is the magnitude of the force vector ${\bf F}$ along the planar
curve, given by
\begin{equation} \label{eq:FPEE}
\mathbf{F} = \left(\frac{\kappa^2}{2} - c \right) \mathbf{T} + \kappa'
\mathbf{N}\,.
\end{equation}
${\bf F}$ is conserved along the curve, i.e., ${\bf F}'=0$.  The quadrature
(\ref{planarquadrature}) is integrable in terms of elliptic functions
\cite{Abramowitz1974, Gradshteyn2007}.  The specific function will
depend on the relative values of $F$ and $c$ \cite{Langer1984,
SingerSantiago2008}.
\vskip1pc \noindent
If $F^2<c^2$ the FS curvature is given by
\begin{equation} \label{kappadn}
 \kappa(s) = 2 \, q \, \dn[q s, u]\,, \quad F = u \, q^2\,, \quad c = q^2 \,
(2 - u)\,,
\end{equation}
where the Jacobi elliptic function with argument $\phi$ and parameter $u$ is
defined by $\dn^2[\phi,u] = 1 -u \, \sn^2 [\phi,u]$ and $\sn [\phi,u]$ is the Jacobi elliptic sine \cite{Abramowitz1974}. In one period the curvature will oscillate asymmetrically about the value $\sqrt{2 c}$ between two positive (or negative) values, so one has $0<\kappa_{\rm min}<\kappa<\kappa_{\rm max}$, where $\kappa_{\rm min} = 2 \, q \, \sqrt{1-u}$ and $\kappa_{\rm max} = 2 \, q$ ($0 \leq u \leq 1$, so $F>0$ and $c>0$). Thus the curve has an {\it orbitlike} shape.
\vskip1pc \noindent
If $F^2 = c^2$ the curvature is given by
\begin{equation} \label{kappasech}
 \kappa(s) = 2 \, q \, \sech \, q s \,, \quad F=c=q^2\,.
\end{equation}
The curve is not periodic; the curvature is positive everywhere but it takes an infinite length for the curve to become planar. That is $\kappa \rightarrow 0$
as $s\rightarrow \pm \infty$. This curve is termed {\it borderline}, corresponding as it does to the separatrix separating orbital and wavelike behavior.
\vskip1pc \noindent
If $F^2 >c^2$ the curvature is given by
\begin{equation} \label{kappacn}
 \kappa (s) = 2 \, \sqrt{u} \, q \, \cn[q s, u] \,, \quad F=q^2\,, \quad c =
q^2(2 \,u -1)\,,
\end{equation}
where $\cn[\phi, u]$ is the Jacobi elliptic cosine \cite{Abramowitz1974}. The curvature oscillates symmetrically about $0$ between $\pm \kappa_M$, i.e., $-\kappa_M < \kappa < \kappa_M$, where $\kappa_M = 2 \,
\sqrt{u}\, q$ ($0 \leq u \leq 1$), so the curve adopts a {\it wavelike} shape.
\vskip1pc \noindent
The parameters $q$ and $u$ are determined by specifying BCs,
which requires  knowledge of the embedding functions of the curve.
\vskip1pc \noindent
The Cartesian coordinates of the curve ${\bf Y}= (x(s),y(s))$ and its tangent
vector ${\bf T}= (x'(s),y'(s))$ can be expressed in terms of the curvature.
Since ${\bf F}$ is a constant vector, it is more convenient to work in
Cartesian coordinates and align it along direction $\hat{\bf x}$, so ${\bf F} = F \, \hat{\bf x}$, with $F$ constant. Projecting ${\bf F}$ onto the Cartesian
basis $(\hat{\bf x}, \hat{\bf y})$, yields
\begin{equation}
x' \left(\frac{\kappa^2}{2} - c\right) - y' \kappa' = F\,, \qquad y'
\left(\frac{\kappa^2}{2} - c\right) + x' \kappa' = 0\,.
\end{equation}
Taking linear combinations of these equations, we get the components of the
tangent vector in terms of the curvature and its derivative,
\begin{equation} \label{factquad}
F \, x' = \frac{\kappa^2}{2} - c \,, \qquad F \, y' = -\kappa' \,.
\end{equation}
This can be regarded as a factorization of the quadrature (\ref{planarquadrature}). Introducing the angle $\psi$ that the curve makes
with the direction  of $\hat{\bf x}$, the components of the tangent vector can
be written as $x'= \cos \psi$ and $y'= \sin \psi$. Also, the curvature is given by the derivative of this angle, $\kappa = \psi'$. Equations (\ref{factquad})  can now be mapped into the statement of  conservation of energy $E$ and the  equation of motion of a pendulum: arclength is identified with  time and
$\psi$ is identified with the angle that the pendulum makes with the vertical
(see, for example, \cite{Djondjorov2008}).
\vskip1pc \noindent
Let arclength be measured from the midpoint of the curve, so $x(0)=0$.  Let the contact points occur at an arclength $s=\pm s_b$.  The normalized loop length
is $l=4 s_b = 2 \pi r$, where $r=R/R_0$. Confinement places two constraints on the curve. First, since the curve must align with the container at points of contact, the tangent vector must be vertical at liftoff, therefore $x'(\pm s_b) = 0$ and $y'(\pm s_b)= \mp 1$. Thus, from Eq. (\ref{factquad}) one has that the constants $c$ and $F$ are given by the curvature and its derivative at the contact points,
\begin{equation} \label{kappabtgvrt}
\kappa(\pm s_b) = \sqrt{2\,c} \quad (c>0)\,, \quad \kappa'(\pm s_b) = \pm F\,.
\end{equation}
The first condition permits one to obtain the angular wavenumber $q$ in terms
of the modulus $m$. Using expressions (\ref{kappadn}), (\ref{kappasech}) and
(\ref{kappacn}) of $\kappa$ and $c$ in terms of $q$ and $u$, this condition
read for each case
\begin{subequations} \label{qsbPEE}
 \begin{align}
 q s_b & = {\rm F}\left[\frac{\pi}{4},u\right] \,, &  F^2 < c^2 \,,\\
 q s_b & = \arcsinh \,1 \,, & F^2 = c^2\,,\\
 q s_b & = {\rm F}\left[\arcsin \frac{1}{\sqrt{2 u}},u\right] \,  & F^2 >
c^2\,,
\end{align}
\end{subequations}
where ${\rm F}[\phi,u]$ is the incomplete elliptic integral of the first kind, \cite{Abramowitz1974}. This condition is of an intrinsic character, for it determines the distance of the contact points along the curve. The second condition due to the confinement on the curve is of extrinsic character, for it fixes the spatial distance between the contact points: the coordinates $x$ of the contact points are constrained to be equal to the cylinder radius,
\begin{equation} \label{xR0const}
x(\pm s_b) = \pm R_0\,.
\end{equation}
To implement this constraint, one needs to integrate Eqs. (\ref{factquad}). The second equation in (\ref{factquad}) can be integrated readily, which is a
consequence of the rotational invariance about the direction orthogonal to the
plane of the curve. Indeed, the integration constant is identified as the
conserved torque along direction $\hat{\bf z}$; that is, $F y = -\kappa - M$,
where $M = {\bf M} \cdot \hat{\bf z}$. By setting the contact points at $y=0$,
it follows that the magnitude of the torque is given by the negative of the
curvature at the contact points, $M = -\kappa_b$. Therefore, the coordinates of the curve are given by
\begin{equation}
x(s) = \frac{1}{F}\displaystyle\int\limits_{0}^s d t
\left(\frac{\kappa(t)^2}{2} - c\right) \,, \quad y(s) =
\frac{1}{F}\left(\kappa_b - \kappa(s)\right) \,.
\end{equation}
In full, the Cartesian coordinates for each case read
\begin{subequations}
 \begin{align}
x &=  \frac{2}{u \, q} \, {\rm E} [\am [q s, u],u] +
\left(1-\frac{2}{u}\right) \, s \,, &y &= \frac{2}{u \,q} \,
\left(\sqrt{1-\frac{u}{2}} - \dn[q s, u]\right) \,, &  F^2 &< c^2\,;
\label{xy1Eplanar}\\
x &= \frac{2}{q} \, \tanh \, q s - s\,,  &y &=  \frac{2}{q} \,
\left(\frac{1}{\sqrt{2}} -\sech \,q s  \right) \,, &  F^2 &= c^2 \,;
\label{xy2Eplanar}\\
x &=\frac{2}{q} \, {\rm E} \left[\am[q s,u],u\right] - s\,, & y & = \frac{2}{q}
\left(\sqrt{u - \frac{1}{2}}  -  \sqrt{u}\, \cn[q s,u]\right) \,, &  F^2 & >
c^2 \,. \label{xy3Eplanar}
 \end{align}
\end{subequations}
where ${\rm E}[\phi,u]$ is the incomplete elliptic integral of the second kind and $\am[\phi,u]$ is the Jacobi amplitude \cite{Abramowitz1974}. This segment forms the bottom half of the free loop; the upper half is obtained by an appropriate up-down reflection.
\vskip1pc \noindent
Imposing condition (\ref{xR0const}) in Eqs. (\ref{xy1Eplanar}) -
(\ref{xy3Eplanar}) provides another expressions of the wavenumber $q$ in terms
of the parameter $m$, obtaining for each case
\begin{subequations} \label{R0qEFPEE}
 \begin{align}
R_0 \, u \, q & = 2 \, {\rm E} \left[\frac{\pi}{4},u\right]- \left(2-u\right)
\,{\rm F} \left[\frac{\pi}{4},u\right] \,, &  F^2 < c^2 \,,\\
R_0 \, q & = \sqrt{2} - \arcsinh \,1 \,, & F^2 = c^2\,,\\
R_0 \, q & = 2 \, {\rm E}\left[\arcsin \frac{1}{\sqrt{2 u}},u\right]- {\rm
F}\left[\arcsin \frac{1}{\sqrt{2 u}},u\right] \,  & F^2 > c^2\,.
\end{align}
\end{subequations}
By combining Eqs. (\ref{qsbPEE}) and (\ref{R0qEFPEE}) and using the relation
$s_b= \pi \,r/2 $ one determines the excess radius $\Delta r = r -1$ as a
function of $u$
\begin{subequations} \label{gammamPEE}
 \begin{align}
\Delta r & = \frac{u \,{\rm F}[\frac{\pi}{4},u]}{\pi \, \left({\rm E}
\left[\frac{\pi}{4},u\right]- \left(1-\frac{u}{2}\right) \,{\rm F}
\left[\frac{\pi}{4},u\right]\right)} -1 \,, &  F^2 < c^2 \,,\\
\Delta r & = \frac{2 \, \arcsinh 1}{ \pi \, \left(\sqrt{2} - \arcsinh
\,1\right)} -1\,, & F^2 = c^2 \,,\\
 \Delta r & = \frac{{\rm F}\left[\arcsin \frac{1}{\sqrt{2 u}},u\right]}{\pi \,
\left({\rm E}\left[\arcsin \frac{1}{\sqrt{2 u}},u\right]-\frac{1}{2} {\rm
F}\left[\arcsin \frac{1}{\sqrt{2 u}},u\right]\right)}-1 \,,  & F^2 > c^2\,.
\end{align}
\end{subequations}
This makes it possible to parametrize the loop by $u$. Alternatively one can
solve Eqs. (\ref{gammamPEE}) numerically in each case to determine $u$ for a
given excess radius $\Delta r$.
\vskip1pc \noindent
The total energy of the loop is easily obtained by noticing from Eq.
(\ref{factquad}) that the energy density is proportional to $x'$. Thus, the
energy of one quarter of the loop is given by
\begin{equation}
 \frac{H_B}{4} = \frac{1}{2}\displaystyle\int\limits_{0}^{s_b} {\rm d}s \,
\kappa^2 = \displaystyle\int\limits_{0}^{s_b} {\rm d}s \, \left( F \, x' +
c\right) = F \, R_0 + c \, s_b\,,
\end{equation}
which establishes the role of the constants $F$ and $c$ as the conjugate
variables to the extrinsic and intrinsic lengths of the curve respectively (cf. Eq. (\ref{hBDeltazphiL}) for the confined case). Using Eqs. (\ref{qsbPEE}) and
(\ref{R0qEFPEE}), one gets that the total energy of the loop in  each of the
three cases by
\begin{subequations}
 \begin{align}
H_{B} & = 8 \, q\, {\rm E}\left[\frac{\pi}{4},u\right]\,,  &  F^2 < c^2 \,,\\
H_{B} & = 4 \, \sqrt{2}\, q \,,  &  F^2 = c^2 \,,\\
H_{B} & = 8 \, q \, \left({\rm E}\left[\arcsin \frac{1}{\sqrt{2
u}},u\right]-\left(1-u \right) {\rm F}\left[\arcsin \frac{1}{\sqrt{2
u}},u\right] \right)  \,,  &  F^2 > c^2 \,.
\end{align}
\end{subequations}
One quarter of the confined loops are plotted for increasing radius in Fig.
\ref{Fig13}(a). If $u=0$ one has a circle. In the range $0 < \Delta r < 0.053$
($0<u<1$) the curve is consists of orbitlike elastica segments, whereas for
$\gamma > 0.053$ ($1/2<u<1$, $u$ decreases as the length increases)  it
consists of wavelike elastica segments (at $\Delta r = 0.053$ ($u=1$) consists
of borderline elastica segments). In this wavelike regime, when $\Delta r =
0.393$ ($u=1/2$) the curve adopts the limit shape shown in Fig. \ref{Fig13}(a)
and by further increasing the length, the middle region of the loop--where the
curve makes contact with the cylinder--will elongate, so the two limit arcs
will be separated by straight lines. The energy of the loops is plotted in Fig. \ref{Fig13}(b) as a function of their radius. The bending energy $H_B$ of the
confined loops begins at $H_{B} / \pi =1/R_0$ (circle with $u=0$) and decreases monotonically as the radius of the loop increases, saturating  when the
limit shape with straight segments is reached  for such lines do not contribute to the bending energy.\footnote{This decay is slower than that of the  energy in a free loop $H_{B0}=\pi/R$,  and achieves a limit value rather than tending to zero for very long loops.}
The normal force acting on the point where contact with the cylinder occurs is
given by $\lambda = 2 \, F$ (each arc contributes with an amount $F$). This can be seen by considering a stress source of magnitude $\lambda$ acting on the
point of contact at $s=s_b$, in the EL equation (\ref{planarELeq}),
\begin{equation}
 \varepsilon_{\bf N} = \kappa'' + \kappa \left(\frac{\kappa^2}{2} - c\right) =
\lambda \, \delta (s-s_b)\,.
\end{equation}
Integrating this equation in the neighborhood of the contact point and taking
into account that the curvature is continuous across the contact
region,\footnote{If an adhesion energy is considered, the curvature would have
a discontinuity \cite{DesMulGuv2007, Seifert1994}.} one has
\begin{equation}
\int_{s_b - \delta s}^{s_b + \delta s} \, {\rm d}s \, \varepsilon_{\bf N} =
\kappa'_+ - \kappa'_- = \lambda \,,
\end{equation}
where $\kappa'_+$ and $\kappa'_-$ are the limits of the arclength derivative of the curvature, on the top and bottom sides of the contact point. Thus, the force exerted on the loop is reflected by the discontinuity in the derivative of its curvature. This is analogous to the force exerted on a fluid vesicle by a ring, presented in Ref. \cite{Bozic2014}. From the BCs. (\ref{kappabtgvrt}),
one has $\kappa'_\pm = \pm F$, so  $\lambda=2 F$.\footnote{The change of sign
between the up and down regions stems from the fact that it is obtained by a
reflection of the top arc, rather than a periodic continuation of the
corresponding elastica.} Moreover, from Eqs. (\ref{kappadn})-(\ref{kappacn})
follows that $\lambda$ is proportional to the squared wavenumber, which is
plotted in Fig. \ref{Fig13}(c). As the loop radius increases, $\lambda$
increases from $0$ (circle) to a constant value $1.435$ (limit arc). When the
straight lines appear, they do not exert a force on the cylinder and the
corresponding normal force for each arc $\lambda = F$ is non vanishing only at the points of detachment.

\section{Closed curves with zero axial force and torque} \label{Sect:f0m0}

\begin{figure}[htb]
\begin{center}
\includegraphics[scale=0.59]{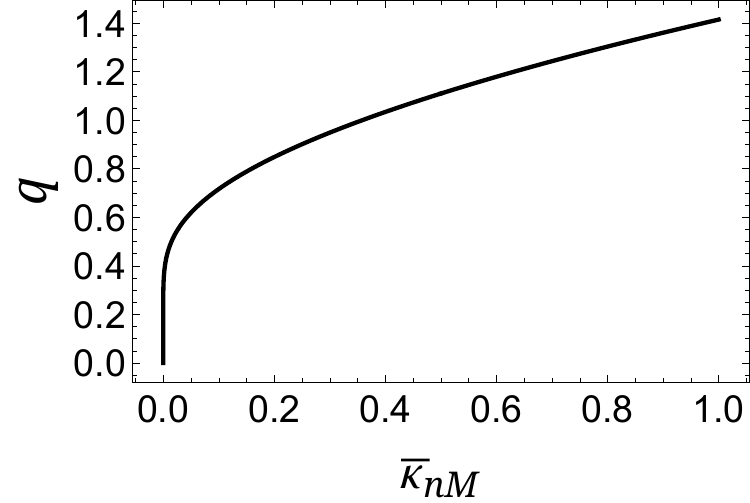}
\caption{Allowed wavenumbers for loops with zero axial torque and force. The
loops are defined for wavenumbers given by rational numbers in the range
$0<n/p<\sqrt{2}$.
} \label{Fig11}
\end{center}
\end{figure}

\begin{figure}[htb]
\begin{center}
  \begin{tabular}{cccc}
  $\vcenter{\hbox{\includegraphics[scale=0.55]{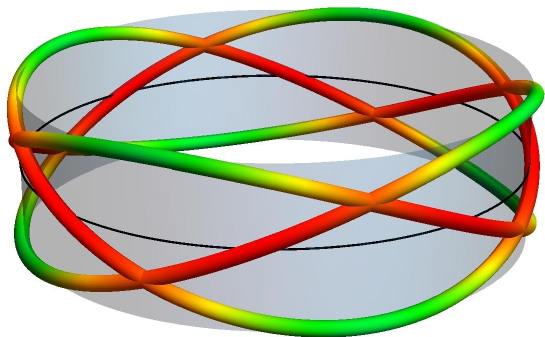}}}$ &
  \quad $\vcenter{\hbox{\includegraphics[scale=0.45]{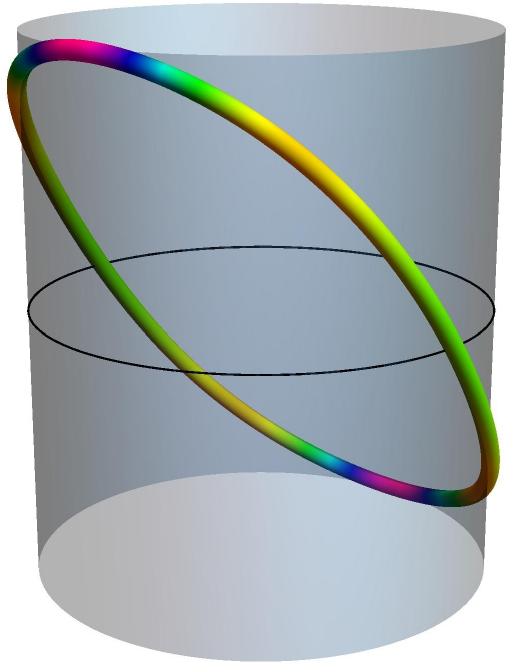}}}$ &
  \quad $\vcenter{\hbox{\includegraphics[scale=0.6]{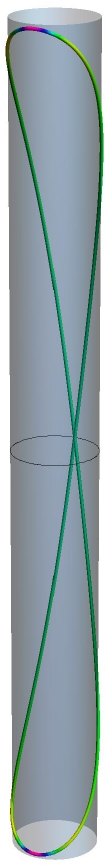}}}$ &
  \quad $\vcenter{\hbox{\includegraphics[scale=1]{Fig5g.pdf}}}$\\
  {\small (a)} &
  \quad {\small (b)} &
  \quad {\small (c)} &
\end{tabular}
\end{center}
\caption{(Color online) Curves with zero axial force and torque with (a) $p=3$, $n=4$ and $\Delta r =  0.122$ ($\alpha_{M} = 0.125 \, \pi$); (b)  $p=1$, $n=1$
and $\Delta r =0.297$ ($\alpha_{M} = 0.296 \pi$); (c) $p=2$, $n=1$ and $\Delta
r =4.8$ ($\alpha_{M} = 0.461 \pi$).  The normalized magnitude of the normal
force $\lambda$ is color-coded in these figures.}
\label{Fig12}
\end{figure}

Loops of special interest are those in which not only the axial force vanishes
but, in addition, the axial torque vanishes, i.e., $m=0$.\footnote{Note, of
course, that the total non-conserved, torque does not vanish in these states.}
While this condition is met in infinitely long loops, rather surprisingly it is also met non-trivially for loops of specific length and precise values of $n$
and $p$. Now the constant $c$ given in Eq. (\ref{calphaMm}) simplifies to  $c = -\cos^4_{M}/2=\bar{\kappa}^2_{n\,M}/2$, so the quadrature (\ref{2ndintcylconf}) reduces to $\alpha'{}^2  = \cos^4 \alpha - \cos^4 \alpha_M$, which can be recast  in terms of the scaled normal curvature as
\begin{equation} \label{knquad}
 \bar{\kappa}_n'=2 \, \sqrt{ \bar{\kappa}_n \,  (1- \bar{\kappa}_n)\,
(\bar{\kappa}_n^2 - \bar{\kappa}_{n M}^2)}\,.
\end{equation}
Solutions exist  in the range  $0\leq \bar{\kappa}_{n M} \leq \bar{\kappa}_n
\leq 1$, so that the normal curvature will oscillate between a minimum
$\bar{\kappa}_{n M} \in (0,1)$ at points where the loop crosses the equator
($\alpha=\alpha_M$) and $1$ at points with minimum and maximum height
($\alpha=0$).  One finds an exact analytic solution upon integrating  Eq.
(\ref{knquad}),
\begin{equation} \label{sknf0m0}
s = \frac{1}{\sqrt{2 \, \bar{\kappa}_{n M}}} \, {\rm F} \left[ \arcsin \,
\sqrt{ \frac{1 -  \bar{\kappa}_{n}}{ 1 - \bar{\kappa}_{n M}} \, \frac{2
\bar{\kappa}_{n M}}{\bar{\kappa}_n + \bar{\kappa}_{n M}}}, \frac{1-
\bar{\kappa}_{n M}}{2}\right]\,.
\end{equation}
Here arc-length $s$ is measured from the tips where $\kappa_n = 1$. The arc-length measured from tip to crossing point is given by $s_M = {\rm K}[(1 - \bar{\kappa}_{n M})/2]/\sqrt{ 2 \, \bar{\kappa}_{n M}}$, where ${\rm K}[u]$ is the complete elliptic integral of the first kind \cite{Abramowitz1974}, so that the  total length of the loop is $l = 4 n s_M$ and the excess radius is $\Delta r = 2 n s_M/\pi -p$.
\vskip1pc \noindent
The normal curvature as a function of arc-length is identified by inverting
(\ref{sknf0m0})
\begin{equation}
 \bar{\kappa}_n (s) = \frac{\bar{\kappa}_{n M} \, {\rm dn}^2 \left[\sqrt{2 \,
\bar{\kappa}_{n M}} \, s , (1 - \bar{\kappa}_{n M})/2 \right]}{1 +
\bar{\kappa}_{n M} - {\rm dn}^2 \left[\sqrt{2 \, \bar{\kappa}_{n M}} \, s , (1
- \bar{\kappa}_{n M})/2 \right]}\,.
\end{equation}
Both the height and azimuthal functions can be determined by combining the quadrature (\ref{knquad}) with the relations $\varphi' = \cos \alpha =
\sqrt{\bar{\kappa}_n}$ and $z' = \sin \alpha = \sqrt{1 - \bar{\kappa}_n}$ and
integrating:
\begin{subequations}
 \begin{eqnarray}
 \varphi(s) &=& \frac{1}{\sqrt{1 + \bar{\kappa}_{n M}}} \, {\rm F}
\left[\arcsin \sqrt{\frac{1 - \bar{\kappa}_{n}(s)}{1 - \bar{\kappa}_{n M}}
}\,,\frac{1 - \bar{\kappa}_{nM}}{1 + \bar{\kappa}_{n M}}\right] \,,\\
 z(s) &=& \frac{1}{\sqrt{2\,\bar{\kappa}_{n M}}} \, \left( {\rm
K}\left[\frac{1}{2}\right] - {\rm F} \left[\arcsin \sqrt{\frac{2 \,
\bar{\kappa}_{n M}}{ \bar{\kappa}_n(s) + \bar{\kappa}_{n M}}
}\,,\frac{1}{2}\right] \right) \,.
\end{eqnarray}
\end{subequations}
The maximum height $z_{\rm max}$ occurs at the tips where $\bar{\kappa}_n =1$.
The total azimuthal extension of the loops is
\begin{equation}
\Delta \varphi = \frac{4 \,n }{\sqrt{1 + \bar{\kappa}_{n M}}} \, {\rm K}
\left[\frac{1 - \bar{\kappa}_{n M}}{1 + \bar{\kappa}_{n M}}\right]\,.
\end{equation}
Closure $\Delta \varphi = 2 \pi p$ implies that the corresponding wave number
is given by
\begin{equation} \label{qm0f0}
q = \frac{n}{p} = \frac{\pi}{2} \, \sqrt{1 + \bar{\kappa}_{n M}} \big{/} {\rm
K} \left[\frac{1-\bar{\kappa}_{n M}}{1 + \bar{\kappa}_{nM}}\right]\,.
\end{equation}
In Fig. \ref{Fig11} $q$ is plotted as a function of $\bar{\kappa}_{nM}$. We see that the wavenumber is bounded in the interval $0 < q < \sqrt{2}$. Excited
states with vanishing torque  do not exist in any sequence with $p=1$; the only such state is an elliptic ground state with $n=1$ and $m=0$ (cf Fig. \ref{Fig3}). For a given $q$ one has to solve numerically Eq. (\ref{qm0f0}) to
determine $\bar{\kappa}_{n M}$. Loops with $q=4/3,1$ and $1/2$ are illustrated
in Fig. \ref{Fig12}. Their corresponding scaled normal curvatures at the
crossing points are $\bar{\kappa}_{n M}=0.853,0.359$ and $0.015$ respectively.

\vskip1pc \noindent
Loops with $\bar{\kappa}_{n M} \approx 1$ will be close to $p$-coverings of the equator. As $\bar{\kappa}_{n M}$ decreases towards zero the loops elevate,
getting more and more vertical as $\bar{\kappa}_{n M}$ tends to zero. In the
limit case $\bar{\kappa}_{n M}=0$, the reduced normal curvature and the
coordinate functions assume a simple form. For $c=0$, the quadrature reduces to $\alpha' = \cos^2 \alpha$, so $\tan \alpha = s$ and $\bar{\kappa}_n = 1/(s^2 +
1)$, whereas the azimuthal and height coordinates for this case can be
represented by $\varphi = \arcsinh s$ and $z =\sqrt{s_M^2 + 1}- \sqrt{s^2 +
1}$, which together permits one to express the normal curvature and the height
in terms of azimuthal angle as $\bar{\kappa}_n = \sech^2 \varphi$ and $z =
\cosh \varphi_M - \cosh \varphi$. Thus it takes an infinite number of windings
and infinite height to go from the tips with $\bar{\kappa}_n = 1$ ($\varphi=0$) to the points crossing the equator with $\bar{\kappa}_n =0$ ($\varphi
\rightarrow \infty$), where the loops have vertical tangents. Therefore the loop has infinite excess radius and winds infinitely many times while completing one period, so $n=1$ and $p \rightarrow \infty$, and consequently $q=0$.
\vskip1pc \noindent
The energy of a curve with length $l = 2 \,s_b$ is given by
\begin{equation}
h_B = 4 \, n \, \displaystyle\int\limits_0^{s_M} {\rm d}s \, \bar{\kappa}^2_n -
 \bar{\kappa}^2_{n M} \, l/2 \,.
\end{equation}
There is no analytic expression in terms of elliptic functions, so it has to be integrated numerically. The energy of the limit loop with $\bar{\kappa}_{n M}=0$ is $h_b = \pi$, ($n=1$).
\vskip1pc \noindent
The magnitude of the normal force is given by
\begin{equation}
\bar{\lambda}(s) = 2 \, \bar{\kappa}_n^2 (s)\, (5 - 6 \, \bar{\kappa}_n(s) ) +  3\, \bar{\kappa}_{n M}^2 \, (2 \, \bar{\kappa}_n (s) - 1)\,.
\end{equation}
For the limit loop with $\bar{\kappa}_{n M}=0$, the magnitude of the normal
force can be parametrized by the azimuthal angle as $\bar{\lambda} = 2
\,\sech^4 \, \varphi \, \left(5 - 6 \,\sech^2 \, \varphi \right)$. It vanishes
as $\varphi \rightarrow \infty$, which corresponds to the points crossing the
equator.

\section{Comparison of cylindrical and planar elastic curves}
\label{App:ComCylPEE}

\begin{figure}[htb]
\begin{center}
  \begin{tabular}{ccc}
  $\vcenter{\hbox{\includegraphics[scale=0.45]{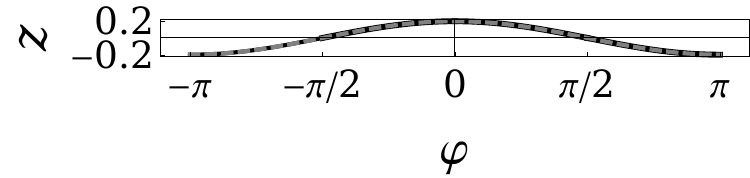}}}$ &
  $\vcenter{\hbox{\includegraphics[scale=0.45]{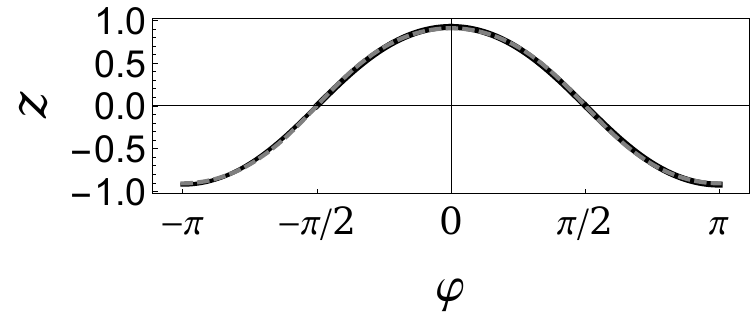}}}$ &
  $\vcenter{\hbox{\includegraphics[scale=0.45]{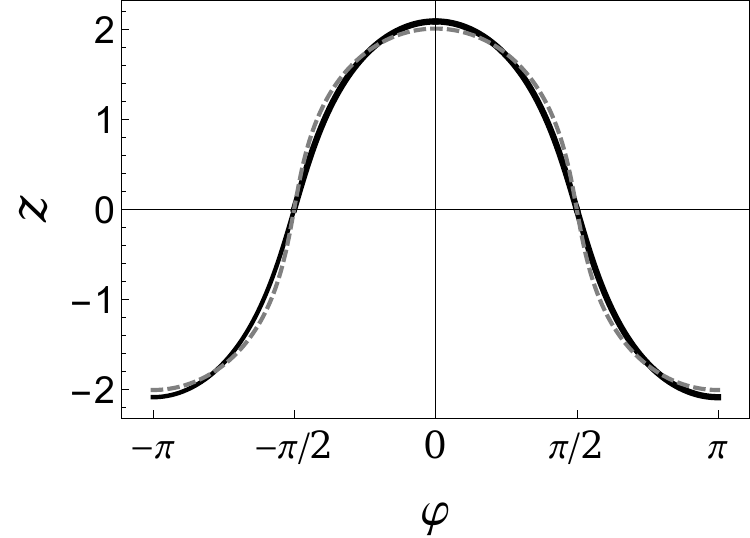}}}$ \\
    (a)  &  {\small (b)} &  {\small (c)} \\
  $\vcenter{\hbox{\includegraphics[scale=0.45]{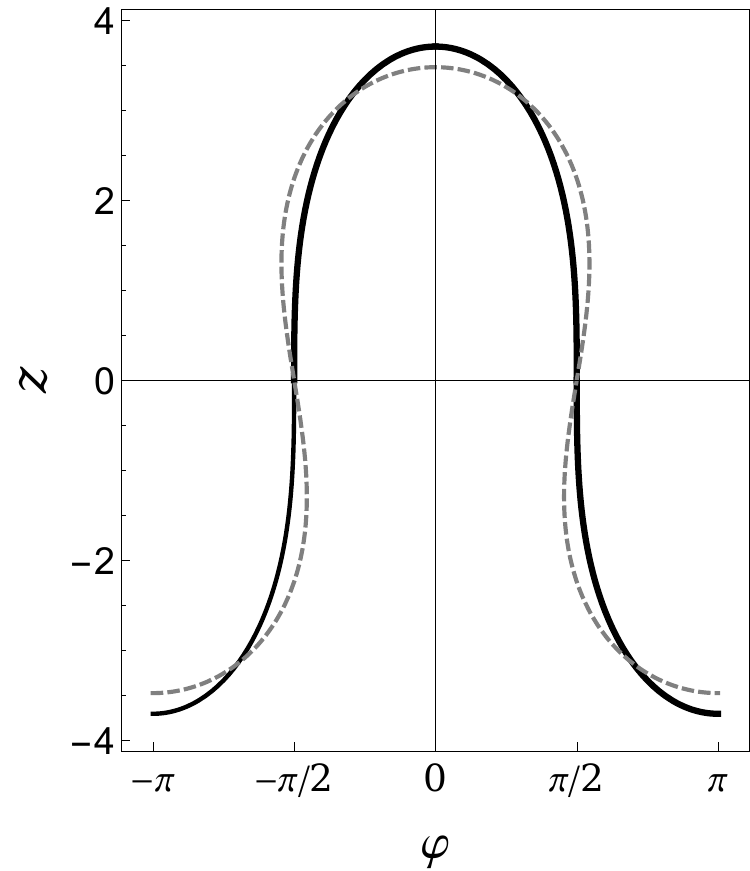}}}$ &
  $\vcenter{\hbox{\includegraphics[scale=0.45]{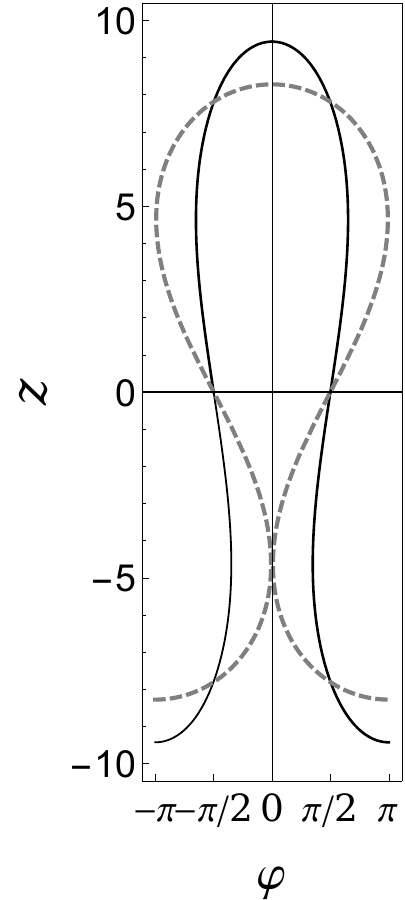}}}$ &
  $\vcenter{\hbox{\includegraphics[scale=0.425]{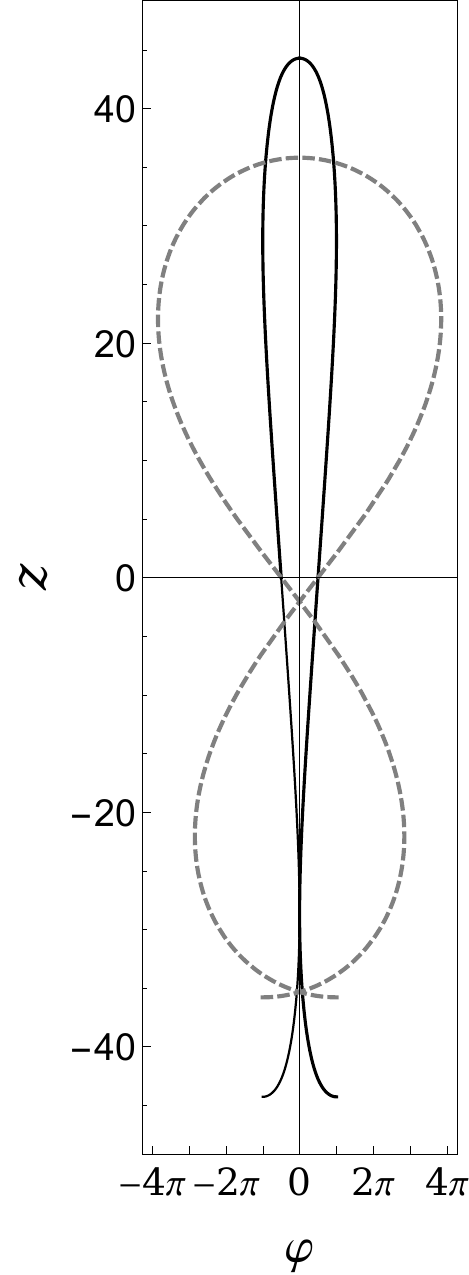}}}$ \\
  {\small (d)} &   {\small (e)} &   {\small (f)}
\end{tabular}
\end{center}
\caption{Unfolding loops in the ${\rm L}_{1,1}$ sequence illustrated in Fig.
\ref{Fig5} are represented by black solid curves, whereas their planar
Euler-elastic counterparts (equal length and same midpoints) are represented by gray dashed curves. For small excess radius the two coincide, but they differ
increasingly as the excess radius grows.} \label{Fig14}
\end{figure}

\begin{figure}[htb]
\begin{center}
  \begin{tabular}{ccc}
  $\vcenter{\hbox{\includegraphics[scale=0.45]{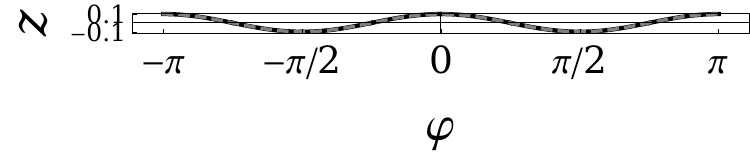}}}$ &
  $\vcenter{\hbox{\includegraphics[scale=0.45]{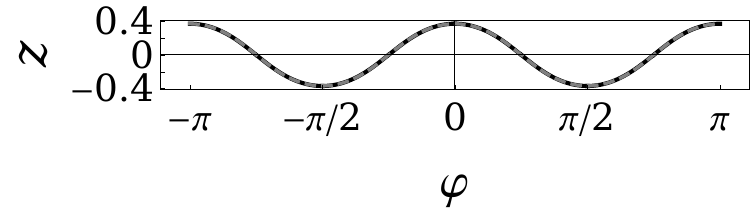}}}$ &
  $\vcenter{\hbox{\includegraphics[scale=0.45]{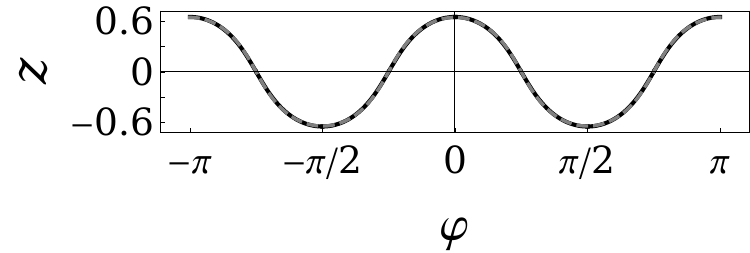}}}$ \\
    (a)  &  {\small (b)} &  {\small (c)} \\
  $\vcenter{\hbox{\includegraphics[scale=0.45]{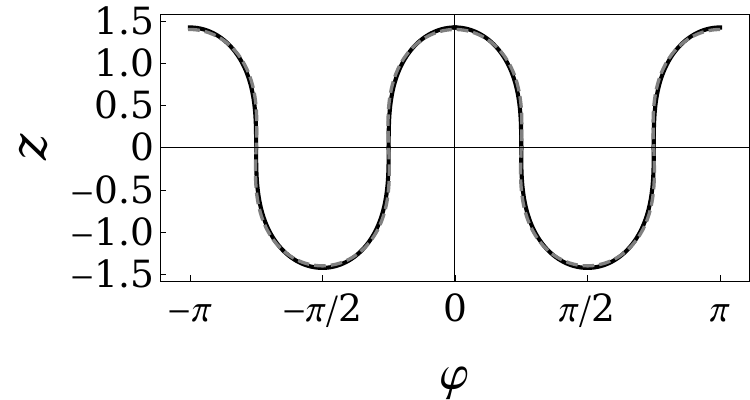}}}$ &
  $\vcenter{\hbox{\includegraphics[scale=0.4]{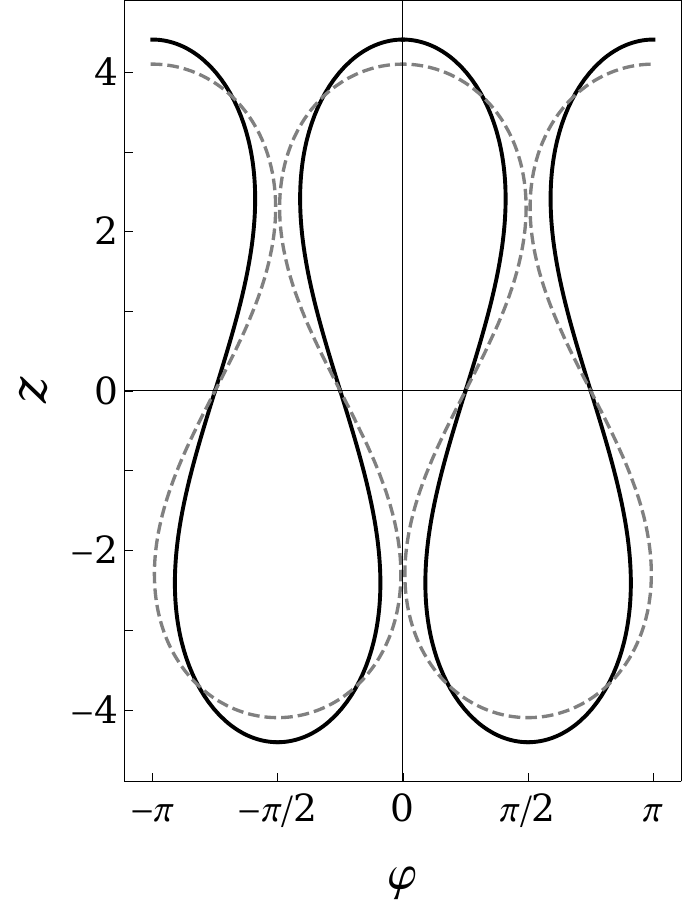}}}$ &
  $\vcenter{\hbox{\includegraphics[scale=0.425]{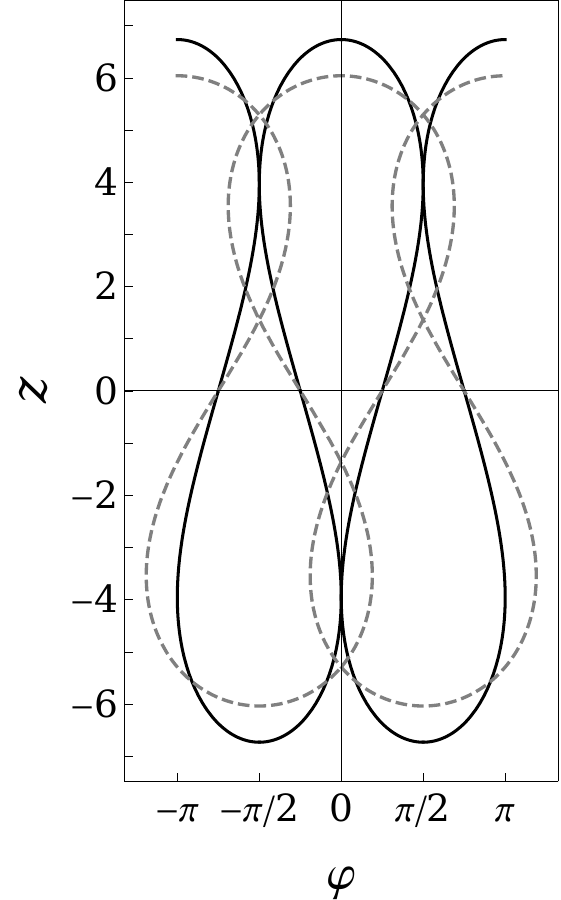}}}$ \\
  {\small (d)} &   {\small (e)} &   {\small (f)}
\end{tabular}
\end{center}
\caption{Unfolding loops in the ${\rm L}_{1,2}$ sequence illustrated in Fig.
\ref{Fig6} are indicated by solid black curves. Their planar counterparts
(equal length and same midpoints) are indicated by gray dashed curves. In
contrast to the $L_{1,1}$ sequence, the two agree pretty well even when the
length is significant.
} \label{Fig15}
\end{figure}

\begin{figure}[htb]
\begin{center}
  \begin{tabular}{ccc}
  $\vcenter{\hbox{\includegraphics[scale=0.35]{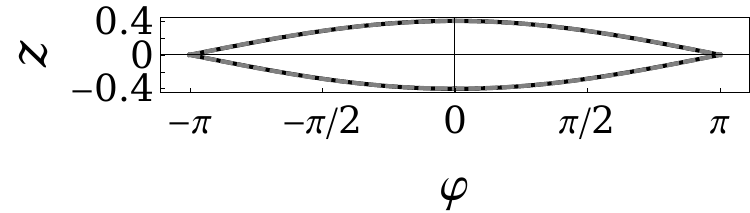}}}$ &
  $\vcenter{\hbox{\includegraphics[scale=0.35]{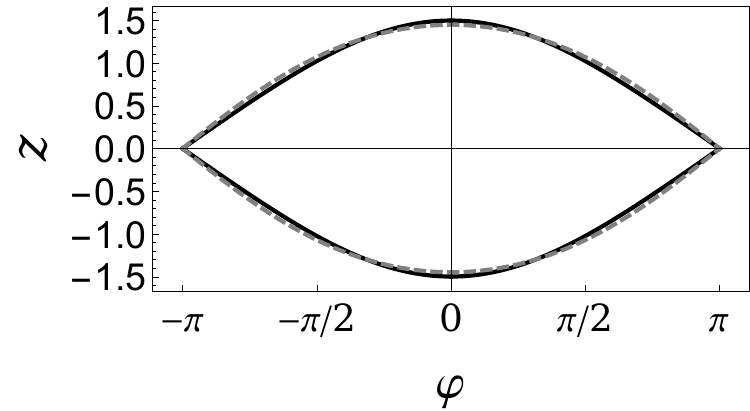}}}$ &
  $\vcenter{\hbox{\includegraphics[scale=0.35]{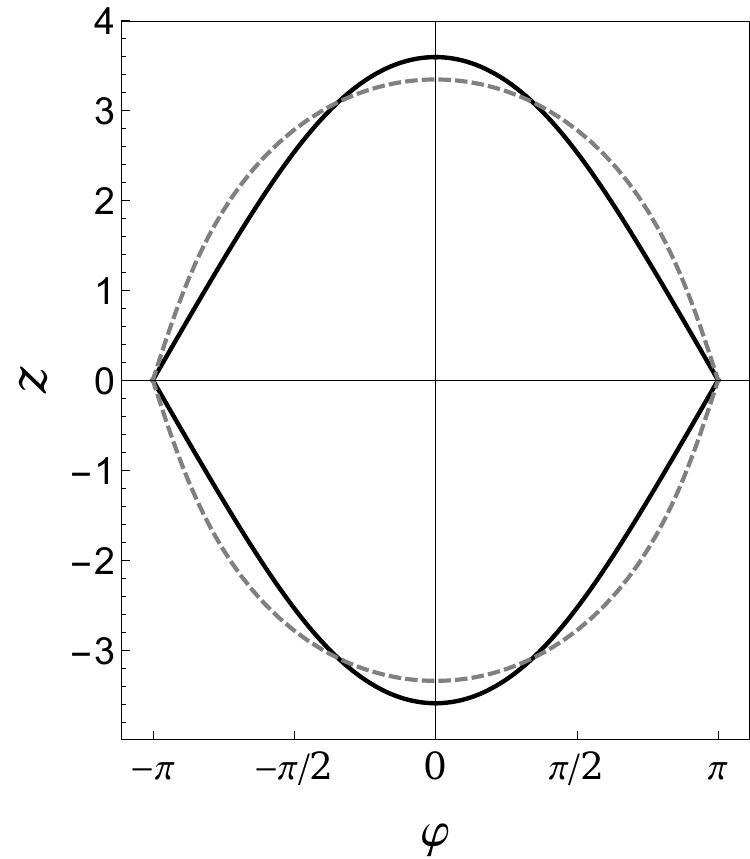}}}$ \\
    (a)  &  {\small (b)} &  {\small (c)} \\
  $\vcenter{\hbox{\includegraphics[scale=0.4]{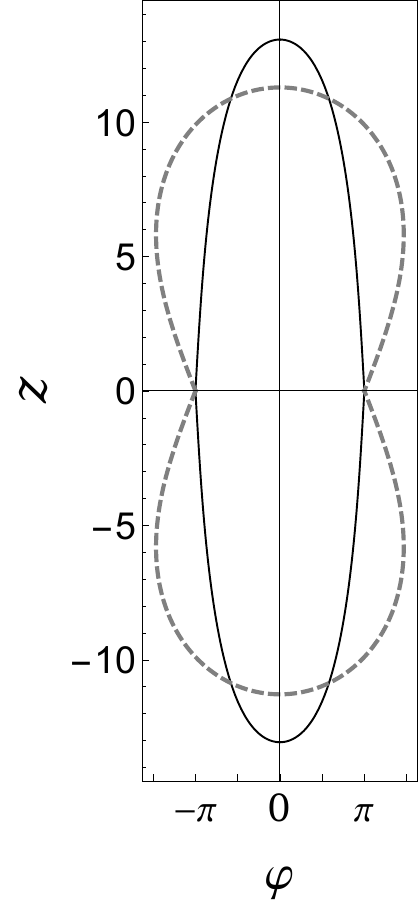}}}$ &
  $\vcenter{\hbox{\includegraphics[scale=0.45]{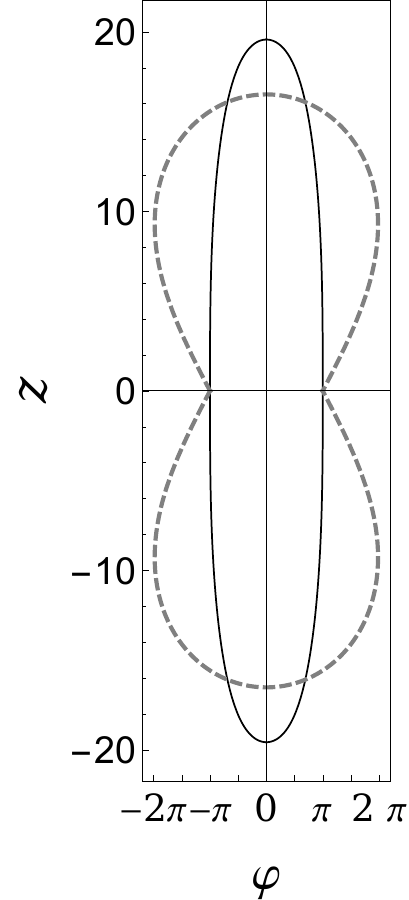}}}$ &
  $\vcenter{\hbox{\includegraphics[scale=0.5]{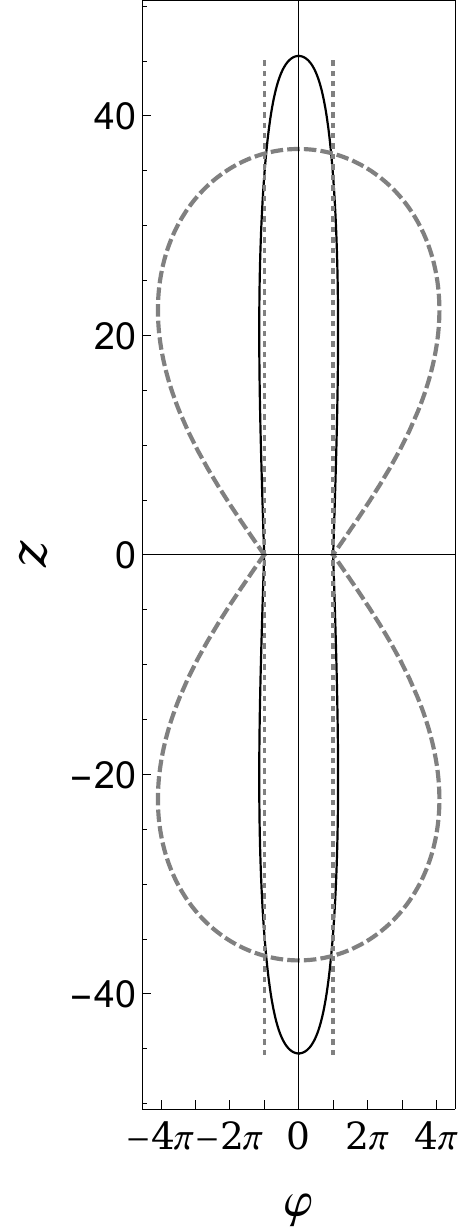}}}$ \\
  {\small (d)} &   {\small (e)} &   {\small (f)}
\end{tabular}
\end{center}
\caption{Unfolded version of loops of sequence ${\rm L}_{2,1}$ illustrated in
Fig. \ref{Fig7} are represented by black solid curves, whereas the planar
Euler-elasticas with the same length and same midpoints are represented by gray dashed curves.} \label{Fig16}
\end{figure}

To visualize the development of overhangs and the subsequent self-intersection
of the loops as the length is increased, it is useful exploit the isometry
between the cylinder and the plane. This is achieved by cutting the cylinder
along the meridian passing through the lower turning point along the loop and
``unrolling'' it, as indicated in Fig. \ref{Fig14}. If $\alpha_M \leq \pi/2$,
the azimuthal extension of the loop on either side of the equator is bounded by $\pi$ ($\Delta \varphi_{max} \leq \pi$); see Figs. \ref{Fig14} (a)-(d).\footnote{The intersections of the loop with the  equator are always
separated by $\pi$.} Overhangs (with $\varphi'=0$),  signaling the appearance
of lobes, first appear on the equator. If $\alpha_M > \pi/2$,  then $\Delta
\varphi_{max} > \pi$, and the loop necessarily  develops overhangs; see Fig.
\ref{Fig14}(e). The azimuthal extension grows monotonically with length: at
some point $\Delta \varphi_{max} = 2 \pi$ and the loop makes self contact for
the first time; see Fig. \ref{Fig14}(f). As the length is increased further,
the overhangs grow indefinitely; overlapping repeatedly as they wrap the
cylinder.
\vskip1pc \noindent
It is also useful to compare loops on the cylinder with the corresponding
planar Euler-elastic curves (dashed gray curves) of the same length passing
through the same two points on the unrolled equator. These are superimposed on
the unrolled loops in Fig. \ref{Fig14}. Initially the cylindrical loop behaves
like a planar elastic curve (Figs. \ref{Fig14}(a)-(b)). As the length is
increased, however, the term in the potential quadratic in the normal curvature begins to play a role and the behavior of the two diverges, as indicated in
Figs. \ref{Fig14}(c)-(d). In a long loop, this term dominates: its effect is to elongate the cylindrical loop along the axial direction. This contrasts
with its rounded planar counterpart which distributes its curvature equally
in azimuthal and vertical directions; see \ref{Fig14}(e)-(f). The corresponding evolutions of the ${\rm L}_{12}$ and ${\rm L}_{21}$ are presented in Figs. \ref{Fig15} and \ref{Fig16}. For the latter sequence, the cut is made at the point of self-intersection on the equator. In Fig. \ref{Fig16}(f) one can see that loops with obtuse angles extend over a azimuthal range  $>2 \pi$ (the vertical dotted gray lines).

\end{appendix}

\bibliography{bibliography}

\end{document}